\documentstyle[prl,aps,twocolumn,epsf]{revtex}

\newcommand{\beq}{\begin{equation}}
\newcommand{\beqn}{\begin{eqnarray}}
\newcommand{\eeq}{\end{equation}}
\newcommand{\eeqn}{\end{eqnarray}}
\newcommand{\vp}{\varphi}

\def\ssec{\subsection}

\newcommand{\ts}{\textstyle}
\def\bfg{\begin{figure}}
\def\efg{\end{figure}}
\def\l{\label}
\def\rd{\displaystyle{\cdot}}
\def\ts{\textstyle}

\begin{document}
\tighten
\draft
\twocolumn[\hsize\textwidth\columnwidth\hsize\csname
@twocolumnfalse\endcsname
\preprint{OUTH-99/,  HUTP-99/; PURCG-99/2
hep-ph/9901319}

\title{
Metric preheating and limitations of linearized gravity
}

\author{
Bruce A.
Bassett$^{\dagger}$
, Fabrizio
Tamburini$^{\P}$, David I. Kaiser$^{*}$ and Roy
Maartens$^{\ddagger}$ 
}
\address{$\dagger$ Department of Theoretical Physics, Oxford
University, Oxford~OX1~3NP, England}
\address{$\P$ Department of Astronomy, University of Padova, Padova,
Italy}
\address{$^*$  Lyman Laboratory of Physics, Harvard University,
Cambridge~MA~02138, USA}
\address{$\ddagger$ School of Computer Science
and Mathematics, Portsmouth University,
Portsmouth~PO1~2EG, England
}

\date{\today}

\maketitle

\begin{abstract} 

During the preheating era after inflation, resonant amplification
of quantum field fluctuations takes place. Recently it has become
clear that this must be accompanied by resonant amplification of
scalar {\em metric} fluctuations, since the two are united by
Einstein's equations. Furthermore, this ``metric preheating"
enhances particle production, and leads to gravitational
rescattering effects even at {\em linear order}. In multi-field
models with strong preheating ($q \gg 1$), metric perturbations are
driven nonlinear, with the strongest amplification typically on
super-Hubble
scales ($k\rightarrow0$). This amplification is causal, being due
to the super-Hubble coherence of the inflaton condensate, and is
accompanied by 
 resonant growth of entropy perturbations. The
amplification
invalidates the use of the linearized Einstein field equations,
irrespective of the amount of fine-tuning of the initial
conditions. This has serious implications on all scales -- from
large-angle cosmic microwave background (CMB) anisotropies to
primordial black holes. We investigate the $(q,k)$ parameter space
in a two-field model, and introduce the time to
nonlinearity, $t_{\rm nl}$, as the timescale for the
breakdown of the linearized Einstein equations. $t_{\rm nl}$ is a
robust indicator of resonance behavior, showing the fine structure
in $q$ and $k$ that one expects from a quasi-Floquet system, and we
argue that $t_{\rm nl}$ is a suitable generalization of the static
Floquet index in an expanding universe. Backreaction effects are
expected to shut down the linear resonances, but cannot remove the
existing amplification, which threatens the viability of strong
preheating when confronted with the CMB.  Mode-mode coupling and
turbulence tend to re-establish scale-invariance, but this process
is limited by causality and for small $k$ the primordial scale
invariance of the spectrum may be destroyed. We discuss ways to escape
the above conclusions, including secondary phases of inflation and
preheating solely to fermions. The exclusion principle constrains the
amplification of metric perturbations significantly. Finally we
rank known classes of inflation from strongest (chaotic and
strongly coupled hybrid inflation) to weakest (hidden sector, warm
inflation), in terms of the distortion of the primordial spectrum
due to these resonances in preheating. 
\end{abstract}
]
\vskip 2pc

\section{Introduction}\l{sec:Intro}

The successes of the inflationary paradigm are  based on
classical perfection and
quantum imperfections. Inflation leads, essentially
because of the no-hair
theorem, to a nearly maximally symmetric, de Sitter-like state.
In contrast, it also leads, via quantum
fluctuations, to small metric perturbations with a nearly scale-invariant
spectrum \cite{infl,MFB}.
The subsequent evolution of these perturbations, in
the absence of entropy perturbations, is extremely simple---controlled by
conserved quantities
\cite{MFB}.
These allow one almost trivially to
transfer the spectrum on super-Hubble scales from the  end of
inflation, typically near the GUT scale $\sim 10^{16}$ GeV,
down to photon decoupling where, in an $\Omega = 1$ universe,
the large-angle anisotropies in the cosmic microwave background (CMB)
are formed, at a scale $\sim 1$ eV---which spans 25 orders of
magnitude in energy.

That such a very simple picture appears consistent with current CMB
and large scale structure data is remarkable. In this paper we
discuss some limitations of this picture and its fragility in the
context of the recent revolution in inflationary reheating known as
preheating. We will show that for a wide range of models the
post-preheating universe must necessarily have followed a much more
complex route than the simple one outlined above.  This more complex
route typically invokes the full nonlinearity of the Einstein field
equations and, in its most extreme guises, appears incompatible with
current observational cosmology. In more moderate form, however, the
preheating route gives new freedom to the inflationary paradigm in
the form of its predictions for the CMB, large scale structure and
primordial black holes. 

Recent study of the reheating era that follows inflation has
highlighted a wide array of interesting new physical processes and
phenomena. These new features all stem from nonperturbative,
nonequilibrium effects in the inflaton's decay.  Whereas earlier
treatments of post-inflation reheating modeled the inflaton as a
collection of independent particles, each decaying into matter fields
perturbatively \cite{oldreh}, these newer analyses focus on
collective, coherent effects which may allow the inflaton condensate
to transfer much of its energy rapidly into a highly non-thermal
distribution of low-momentum particles.  The inflaton's decay, in
other words, may be {\em resonant}, coherent and dominated by the
peculiarities of quantum statistics
\cite{preh1,num,KLS2,DBDK,prerev}.\footnote{Loosely speaking,
preheating is to the old theory of reheating what the high-powered
laser is to the household light-bulb.} During preheating, certain
Fourier-modes of the matter-field fluctuations grow essentially
exponentially with time, $\delta \varphi_k \propto \exp(\mu_k t)$,
where $\mu_k$ is known as the Floquet index or characteristic
(Lyapunov) exponent. Such rapid growth of the field fluctuations
leads to particle production, with the number density of particles
per mode $k$ growing roughly as $n_k \propto \exp (2 \mu_k t)$
\cite{preh1,KLS2,prerev}. The period of resonant inflaton decay has
been dubbed \lq\lq preheating,"  in order to distinguish it from
earlier perturbative studies of reheating. 

Details of this resonance, such as the rate of particle production
and the momenta of particles produced, remain model-dependent.  Even so,
certain consequences of such exponential growth in the occupation numbers
of low-momentum quanta may be investigated.  Such effects include the
possibility of non-thermal symmetry restoration and defect production
\cite{ntsr,ntsrdefects}, supersymmetry breaking and GUT-scale
baryogenesis \cite{gut}. These kinds of processes follow from the rapid
amplification of matter-field fluctuations, driven during the preheating
era by the oscillation of the coherent inflaton condensate.  The
strongly nonperturbative behavior within the matter-fields sector points
to important cosmological consequences.

In a recent paper \cite{BKM1} (hereafter Paper I), we highlighted a
distinct consequence which {\em must} in general arise during resonant
post-inflation reheating:
{\em the resonant amplification of metric perturbations.}
Essentially, this arises because Einstein's equations directly
couple the explosive growth in the field fluctuations
to the fluctuations in the gravitational curvature, so that
the previous approach of neglecting metric perturbations
is inconsistent.
Unlike some of the possible effects of preheating for the
matter-fields
sector, the rapid growth of metric perturbations might leave distinct,
observable signals in the
CMB radiation, and could dramatically alter primordial black hole
formation \cite{BKM1}.

Here we develop the results of Paper I, via additional qualitative
arguments and extensive numerical
simulations of a two-field model.  These simulations allow
us to extend our discussion of the highly effective {\em symbiotic}
interaction between metric perturbations and particle production during
preheating.  The source of the energy density for both metric
perturbations and particle production is the oscillating inflaton
condensate.  Because
the energy density of gravitational perturbations can carry an
opposite sign to the energy density of the matter-field fluctuations, the
two kinds of perturbations need not compete for the decaying inflaton
energy, but rather can grow resonantly together, coupled by the Einstein
equations \cite{BKM1}.  (Note that we consider only scalar modes of the
metric perturbations here.  An interesting extension would be to consider
effects on vector or vorticity modes of the perturbed metric during
preheating.)

To emphasize this, contrast the resonant evolution of
metric perturbations at preheating with that in traditional studies of
particle production in curved spacetime \cite{BD82}. In the latter case,
particle production arises via the polarization of the vacuum due to the
shear of the metric and its perturbations, and hence energy
conservation leads to a {\em
reduction} in the anisotropy and inhomogeneity of spacetime.\footnote{
The most powerful use of this
idea was in the Misner ``chaotic cosmology" program \cite{misner} whereby
it was hoped that arbitrary initial anisotropy and inhomogeneity might be
smoothed out by particle production \cite{grib}.
}
Instead, the resonant amplification of metric and field fluctuations
described here takes energy from the oscillating {\em isotropic and
homogeneous} part of the gravitational field and so is qualitatively very
different: both metric and field fluctuations can grow simultaneously.

In earlier studies of preheating, which neglected the presence and growth
of metric perturbations, important distinctions were found between
preheating in single-field models as opposed to models with distinct fields
coupled to the inflaton \cite{preh1,KLS2}. As highlighted in Paper I, and
developed further here, it is crucial to attend to these
distinctions when studying the amplification of metric perturbations at
preheating.  When resonant decay is possible at all in the single-field
cases, it is typically rather weak, occurring only within a narrow momentum
range and governed by a small characteristic exponent $\mu_k$.  When
distinct scalar fields are coupled to the inflaton, however, qualitatively
different behavior results throughout most of parameter space:
resonances may be very broad in momentum space, and the rate
of particle production may be much greater than in the single-field cases.
Whereas the expansion of the universe during preheating is often
sufficient to damp dramatically the resonant particle production in the
single-field, narrow-resonance cases, broad resonances in the multi-field
cases often survive in the expanding universe.

As we demonstrate here, broad resonances
in multi-field models can drive the exponential amplification of
metric perturbations, and in these models, analyses based on Floquet
theory remain useful, even when the expansion of the universe is
included. (See Fig. (\ref{fig:p2f1}).)  In general, the effective
characteristic
exponents
$\mu_{k,\rm eff}$ for the growth of metric fluctuations
are greater than or equal to the exponents governing the field
fluctuations' growth.

\begin{figure}
\epsfxsize=3.4in
\epsffile{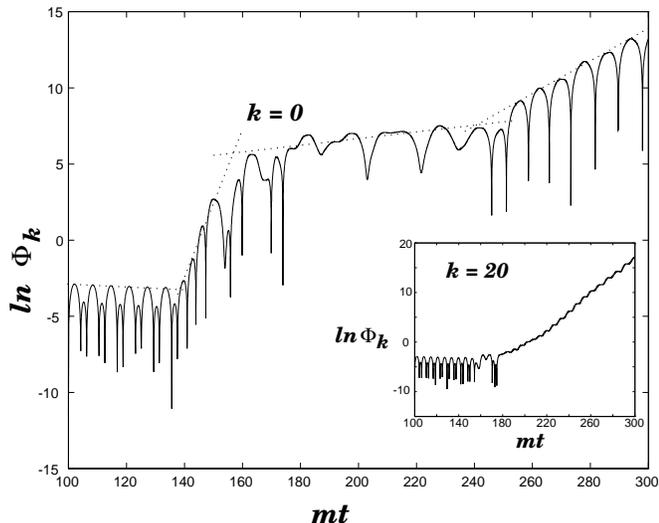}
\caption{Growth of $\ln \Phi_k$ versus $mt$, for $q = 8000$ with
$\Phi \sim 10^{-3}$ initially. $\Phi$ is 
the gauge-invariant gravitational potential, $m$ is the inflaton mass, and
$q$ is the resonance parameter, as defined in Sec. \ref{sec:Preh}.  The 
specific model considered here is given by Eq. (\ref{Vpc}).  The main
plot is for $k = 0$, and the inset shows growth for $k = 20$; the former
is a super-Hubble mode, while the latter is within the Hubble radius.
This plot comes from linearized perturbation theory, and so cannot be used
to describe the modes' behavior in the nonlinear regime.  Note, however,  
the strong breaking of
scale-invariance in the earliest, linear-regime phase:  the two modes go
nonlinear at very different times.}
\l{fig:p2f1}
\end{figure}

Furthermore, as is true of the nonthermal spectra of matter-field
fluctuations amplified during preheating, the strongest resonances
(largest $\mu_k$) for the metric perturbations occur for long-wavelength
modes. Often these modes have wavelengths larger than the Hubble scale at
preheating, $k/a < H$, where $a(t)$ is the cosmic scale factor.  As
discussed in Paper I, such long-wavelength modes may indeed be amplified
causally at preheating; in general, this need not be surprising, since
Einstein's field equations necessarily incorporate relativistic
causality. {\em The resonant
growth of super-Hubble fluctuations derives from the coherence
of the inflaton condensate on these large scales immediately after
inflation, itself a natural consequence of inflation.}
This growth bypasses the standard conservation law since
it is accompanied by {\em resonant growth of entropy perturbations},
arising from the violent transfer of energy from the inflaton.
Earlier studies of metric perturbations induced
during (rather than after) inflation in multi-field models of inflation
indicated that super-Hubble metric perturbations will not, in
general, remain \lq frozen' at their horizon-crossing amplitudes, but will
rather evolve non-adiabatically \cite{multi}. The preheating, exponential
amplification of super-Hubble metric perturbations studied here
represents an extreme example of such dynamic,
non-adiabatic, long-wavelength growth.

By amplifying a distinct, highly scale-dependent spectrum of
metric perturbations after the end of inflation,
multi-field preheating could dramatically alter the observational
consequences often assumed to hold for generic models of inflation. Rather
than being sensitive only to the primordial, \lq stretched' quantum
fluctuations of the inflaton from deep within the Hubble radius during
inflation, the power spectrum of density perturbations created
during inflation would be deformed by the later preheating
dynamics; the spectrum measured by COBE would 
therefore be sensitive only to these combined effects.

Given the robustness of the amplification of long-wavelength metric
perturbations during strong preheating, then, one of the greatest
outstanding
issues regarding the post-inflationary universe is the nonlinear evolution
of these perturbations.  The numerical simulations presented here stem
from integrating the equations of linearized perturbation theory.  The
time it takes for this linearized approximation to break down may be
measured self-consistently, and is discussed below.  Typically, modes go
nonlinear even before the matter-field fluctuations do, and hence well
before what had previously been assumed to be the end of preheating.
Beyond linear order,
mode-mode coupling between metric perturbations of different wavelengths
is likely to be crucial for understanding the evolution of the power
spectrum.  These nonlinear effects would in general remove the individual
\lq fingerprints' of specific inflationary models.  

Naturally, the challenge remains to study this fully nonlinear regime in
quantitative detail.  In preparation for such future studies, we indicate
analytically below some effects which may be expected from such
nonlinearities.
We emphasize that these effects are qualitatively new. The old theory of
perturbation evolution based on conserved quantities \cite{MFB}
allowed for large amplification of metric perturbations at reheating, but
required no production of entropy perturbations, i.e. purely adiabatic
spectra 
on large scales. Perhaps more important, the conserved quantities
were used to
find the metric potential $\Phi$ during inflation by using the
COBE data $\Delta
T/T \sim 10^{-5}$ and the Sachs-Wolfe relation $\Delta
T/T \approx {1\over3}\Phi$. In the usual account, the
large amplification of the $k \rightarrow 0$ modes of $\Phi$ between
inflation
and matter domination then requires fine-tuning of the inflaton mass
($m \sim 10^{-6} M_{\rm pl}$) and self-coupling ($\lambda \sim 10^{-12}$).

What we discuss departs from this in two important ways. First, entropy
perturbations are resonantly amplified during preheating, so that the
standard Bardeen parameter $\zeta$ is not conserved and $\Phi$
is further amplified.  Second, and perhaps more important, we
cannot use the total amplification naively to fine-tune the initial
amplitude of $\Phi$ during inflation to match with the $\Phi \sim 10^{-5}$
at decoupling required by the CMB.  This second point is discussed further
below, in Sec. \ref{sec:Back}.

Work on the evolution of scalar metric perturbations through
preheating began with study of the $k = 0$ mode in simple 
single-field models \cite{KH1,NT1}, for which no super-Hubble 
resonance
was
found. At this early stage, a major concern was the periodically
singular nature of the evolution equations of the Bardeen potential
$\Phi$ due to the appearance of $\dot{\varphi}^{-1}$ terms, where
$\varphi$ is
the oscillating inflaton background field. In contrast,
the evolution of gravitational waves (tensor perturbations) has none
of these problems \cite{gw,geom}.

Historically,  two-field models were
studied soon afterwards \cite{KH2,NT2}. In \cite{KH2}, the
periodic singularities were removed by employing space-time
averaging. This also removed the source of the parametric resonance
and hence no parametric amplification was found. In \cite{NT2}, the
oscillations of the inflaton were retained and the singularities
avoided by using the so-called Mukhanov variables. A resonance in the
scalar metric perturbations was found. However that study was
essentially restricted to classical mechanics -- only the two background
scalar fields were considered and no reheating to non-condensate
modes ($k > 0$) was taken into account. Further, the study was limited to
the first (and
narrowest) resonance band of the Mathieu equation along the line $A
= 2q$. Partly because of this, the authors never
considered what happens when metric perturbations go nonlinear.

Paper I \cite{BKM1} was a response to the perceived limitations of
these earlier works in regard to a realistic treatment of preheating.
Work subsequent to Paper I has re-examined the single-field case
\cite{FB,taru98} and begun study of the complex numerics of the
problem when the metric perturbations become fully nonlinear
\cite{PE}. The need for such a program of study to understand
strong preheating is one of the broad theses of this paper.

The remainder of this paper is organized as follows.  In Section
\ref{sec:Preh}, we
review some basic features of preheating when the coupled metric
perturbations are neglected altogether, and highlight the qualitative, as
well as quantitative, differences between preheating with a single
inflaton field, versus preheating when the inflaton is coupled to distinct
scalar fields.
Section \ref{sec:Concept} presents a sketch of the important conceptual
foundations of
the alternative picture of preheating introduced in Paper I and developed
further here.  Section \ref{sec:MPerts} begins the detailed study of the
coupled metric perturbations.  We present the gauge-invariant
coupled equations for fluctuations of the metric and fields, and
discuss the issues of anisotropic stresses and entropy
perturbations.  In Section \ref{sec:Causal} we address the
question of causality and the amplification of super-Hubble modes. 

In Section \ref{sec:Num},
we focus on a simple two-field model, performing a wide range of
numerical simulations to show the growth and evolution of
fluctuations, and the scale-dependence of amplification.  The time $t_{\rm
nl}$ during which the linearized
approximation remains adequate for the resonantly-amplified metric
perturbations is studied in detail.
Section \ref{sec:Back} turns to the essential role played by nonlinear
features in understanding the preheating epoch, such as backreaction,
mode-mode coupling, and turbulence.  We develop some of the
consequences such strong nonlinearities may have for the perturbation
power spectrum, and discuss possible observational
consequences of the rapid amplification of long-wavelength metric
perturbations.  Section \ref{sec:waysout} discusses possible \lq\lq escape
routes" from the resonant amplification found in our simple model.  In
Section \ref{sec:Hierarchy}, we describe
the hierarchy of inflationary models delineated by the strength of the
amplification of super-Hubble metric perturbations.
Conclusions follow in Section \ref{sec:Conc}.

\section{Preheating Basics} \l{sec:Preh}

\ssec{Minkowski spacetime}

In this section, we review briefly some of the key results for preheating,
as derived in the absence of metric perturbations.  This section is meant
to introduce key ideas and fix our notation; more complete reviews of
preheating may be found in \cite{prerev}.  We begin the discussion in
Minkowski spacetime, and consider after that important deviations from
these results for models in an expanding (but unperturbed)
spacetime.

The equation of motion
for any scalar field $\chi$ coupled via a $g\phi^2 \chi^2$ term to the
inflaton, or for example,  inflaton fluctuations in a model with a
quartic self-coupling, takes the general form in Minkowski spacetime:
\beqn
\nonumber \left[ \frac{d^2}{dt^2} + \Omega_k^2 (t) \right] \chi_k = 0 , \\
\Omega_k^2 (t) \equiv \omega_k^2 + g \varphi_0^2 f (\omega
t) + \lambda_\chi \Sigma (t) .
\label{eom1}
\eeqn
Here $\omega_k^2 = k^2 + m_\chi^2$ is the square of the time-independent
natural frequency of the mode $\chi_k$ in the absence of interactions;
$\varphi_0$ and $\omega$ are the amplitude and frequency of the inflaton's
oscillations, respectively, and $f
(\omega t) \equiv \phi^2 (t) / \varphi_0^2$.  The modes $\chi_k$ thus
evolve with a  time-dependent frequency, $\Omega_k (t)$.

The term $\Sigma (t)$
measures backreaction on the modes $\chi_k$ due to nonlinear $\chi$
self-couplings, such as $\lambda_\chi \chi^4$.  A similar term will
appear in the equation of motion for the oscillating inflaton field,
which encodes the drain of energy from the oscillating inflaton   to
the amplified fluctuations.  It is common in the preheating
literature to treat such nonlinear terms with
either the Hartree-Fock or large-$N$ approximations, in which case $\Sigma
(t) \sim \int d^3k \vert \chi_k \vert^2$.  As discussed below in Sec.
\ref{sec:Back}, these approximations 
capture some elements of the nonlinear structure of the quantum theory,
but neglect such effects as rescatterings between $\chi_k$ modes with
different momenta $k$. These rescattering terms can in many cases become
quite important in the later stages of preheating. In any case,
$\lambda_\chi \Sigma (t) \ll g \varphi_0^2$ for early times, and this
term may be neglected for the first stage of preheating.

Most models of inflation end with the inflaton $\phi$ oscillating
coherently.  These include the common examples of chaotic inflation in
convex potentials, and new inflation with its second-order phase
transition, but excludes models with first-order phase transitions which
complete via bubble nucleation, warm inflation, and non-oscillating
models.  In those cases for which the inflaton does oscillate coherently
at the end of inflation, resonance effects may become crucial.  For early
times after
the start of the inflaton's oscillations, when $\lambda_\chi \Sigma (t)$
is negligible, $f (\omega t)$ is {\em periodic} when studying preheating
in Minkowski spacetime. It is convenient to
write
\beq
\Omega_k^2 (t) = \omega^2 \left[ A_k + 2q f(\omega t) \right] ,
\label{Akq}
\eeq
where $A_k \propto \omega_k^2 / \omega^2$, and $q \propto g \varphi_0^2 /
\omega^2$.  Exact relations depend on specifics of the inflaton
potential $V (\phi)$; for a massive inflaton without self-couplings, $q =
g \varphi_0^2 / (4 m^2)$, where $m$ is the inflaton mass.  (The
factor of 4 in the denominator for $q$ is a convention which comes when we
write the equation of motion for $\chi_k$ in the form of the Mathieu
equation.)  In this early-time regime,
the equation of motion for the modes $\chi_k$ is a linear second-order
differential equation with periodic coefficients. Floquet's theorem then
shows that solutions take the form \cite{floquet}
\beq
\chi_k (t) = P_k (t) e^{\mu_k t} ,
\eeq
where $P_k (t)$ is periodic.  For certain values of $k$,
lying within
discrete \lq resonance bands' $\Delta k$, the Floquet index
$\mu_k$ will have a nonzero real part, indicating an exponential
instability.  Modes within these resonance bands thus grow exponentially
in time.  For the remainder of this paper,
we will write $\mu_k$ as the real part of the Floquet index only.

Resonantly-amplified modes correspond to rapid particle production.  The
occupation number-density for quanta of momentum $k$ may be written simply
in the form
\beq
n_k = \frac{\Omega_k}{2} \left( \vert \chi_k \vert^2 +
\frac{1}{\Omega_k^2} \vert \dot{\chi_k} \vert^2 \right) - \frac{1}{2} .
\label{nk}
\eeq
This follows directly from standard Bogoliubov transformation
techniques \cite{BD82}; its form may be understood intuitively, as
discussed in \cite{KLS2}, as $n_k = \rho_k (\chi) / \Omega_k$, where
$\rho_k(\chi) = {1 \over 2}( \vert \dot{\chi}_k \vert^2 +
\Omega_k^2  \vert \chi_k \vert^2)$ is the energy density carried
by a given $\chi_k$ mode.  From Eq. (\ref{nk}), it is clear
that modes within a resonance band, growing as $\chi_k \sim \exp ( \mu_k
t)$, correspond to a number density of produced particles for that
momentum of $n_k \sim \exp (2 \mu_k t)$.  This is how preheating
produces a highly non-thermal spectrum of particles immediately after
inflation:  only particles in certain discrete bands $\Delta k$ get
produced resonantly, resulting in a highly non-Planckian distribution.
Occupation numbers for quanta lying within the resonance bands grow
quickly,  and the small $k$ modes may well be described as
classical before the end of preheating, in the sense that the density
matrix takes a diagonal form with rapidly oscillating off-diagonal
terms which time-average (in Minkowski spacetime) to zero and are
suppressed in the long-wavelength limit \cite{q2c,son}.  

The strength ($\mu_k$) and width ($\Delta k$) of the resonance depend
sensitively upon the \lq resonance parameter' $q$.  In the narrow
resonance regime, $q \leq 1$, analytic results yield $\mu_k \sim
\Delta k \sim q \omega$, indicating \lq mild' exponential growth
within narrow resonance bands \cite{preh1,KLS2,DBDK}. In the broad
resonance
regime, $q \gg 1$, analytic study of the coupled modes is far more
involved, though significant progress has been made; in this case, the
largest exponents satisfy
$\mu_k \geq O(\omega)$, and the widths of the resonances
scale as $\Delta k \propto q^{1/4}$ \cite{preh1,num,KLS2,NCI}. In all
known cases, the strongest resonances (largest $\mu_k$) occur on
average for low-momentum (small-$k$) modes.

Still stronger resonances occur in models which possess a \lq negative
coupling instability' \cite{NCI}. This instability affects multi-field
models for which the coupling $g\phi^2\chi^2$
between the inflaton and a distinct scalar
field $\chi$ is negative: $g < 0$.  (One can still
maintain a
positive-definite energy density for such models, and hence a stable
vacuum structure, by adding appropriate values of quartic self-couplings
for both $\phi$ and $\chi$.)  Now the
small-$k$ modes $\chi_k$ \lq feel' an {\em inverted} harmonic oscillator
potential, that is, the parameters of Eq. (\ref{Akq}) become approximately
\beq
A_k \approx \frac{k^2 + m_\chi^2}{m^2} + 2q \>,\>\> q \approx \frac{g
\varphi_0^2}{4 m^2},
\eeq
so that $A_k \leq 0$ for $k^2 + m_\chi^2 \leq {1\over2}\vert g \vert
\varphi_0^2$.
Such models lead to highly efficient amplification of these modes $\chi_k$,
now within even wider resonance bands than in the \lq usual'
broad-resonance cases.  For modes subject to a negative coupling
instability, the resonance band widths scale as $\Delta k \sim
\vert q \vert^{1/2}$ rather than as $q^{1/4}$ \cite{NCI}.

In Minkowski spacetime, the resonances which characterize the
earliest stages of preheating will end when the backreaction
of produced particles, $\Sigma (t)$, significantly damps the
inflaton's oscillations, ending the parametric resonance. This
process is typically non-equilibrium and non-Markovian, i.e. it is
very badly modeled by a $\Gamma\dot{\phi}$ term in the equation of
motion for $\phi$ \cite{DBDK,KLS2}. One may find this time
analytically, setting $\lambda_\chi \Sigma (t_{\rm end}) = g \varphi_0^2$
and solving, in the Hartree approximation,  for $t_{\rm end}$ which
yields $t_{\rm end} \sim (4\mu_{\rm max} m)^{-1}\ln
(10^{12}m g^{-5}M_{\rm pl}^{-1})$, where $\mu_{\rm max}$ is the maximum
value 
of $\mu_k$ \cite{KLS2,DBDK}. After $t_{\rm end}$ the
system of coupled oscillators enters the fully non-linear regime; even
here, certain modes $\chi_k$ may continue to grow due to strong
rescattering effects \cite{num,KLS2,DBasym}.

Before considering the important changes to these results in an expanding,
dynamical background spacetime, we first contrast preheating in a
single-field model with preheating in multi-field models.  Consider a
single-field inflationary potential of the form
\beq
V (\phi) = {\ts{1\over2}}{m^2} \phi^2 + {\ts{1\over4}}{\lambda} \phi^4 .
\eeq
Considered in Minkowski spacetime, this model can produce no inflaton
quanta at all when $\lambda = 0$:  fluctuations $\delta \varphi$ obey the
same equation of motion as the
background field $\phi$, and simply oscillate.  When $\lambda \neq 0$ and
$m \neq 0$, the resonance parameter for the fluctuations $\delta \varphi$
takes the form
\beq
q \approx \frac{3 \lambda \varphi_0^2}{4 m^2} .
\eeq
Yet according to the usual treatment of the power spectrum of primordial
density perturbations generated during inflation \cite{infl,MFB}, both
$\lambda$ and $m$ for this model are constrained
by observations of the CMB radiation to take the values $\lambda
\sim 10^{-12}$ and $m \sim 10^{-6} M_{\rm pl}$, where $M_{\rm pl} = 1.22
\times 10^{19}$ GeV is the Planck mass.  Furthermore, for chaotic
inflation, the slow-roll phase will end, and the inflaton will begin to
oscillate, at $\varphi_0 \sim 0.3 M_{\rm pl}$.  These yield $q < 0.1$.
When $m = 0$, the inflaton's frequency of oscillation is well-approximated
by $\omega \approx 0.85 \sqrt{\lambda} \varphi_0$
\cite{KLS2,DBDK}, yielding $q \approx
1$.\footnote{When the inflaton potential contains quartic
self-couplings, the specific resonance structure for coupled fields is
quite different from the case when
$\lambda = 0$; in particular, whereas an infinite hierarchy of resonance
bands results when $\lambda = 0$, only one single resonance band exists
for $\lambda \neq 0, m = 0$.\cite{KLS2,DBDK}  Even so, in {\em both}
cases, preheating in a single-field model is restricted to the
narrow-resonance regime, for
which $q \leq 1$ and $\mu_k \sim q \omega$, and thus these fine-structure
points are not relevant to our present discussion.}  Needless to say,
smaller values of the inflaton's amplitude, corresponding for example
to GUT-scale inflation, will of course yield far smaller values of
$q$.  Thus the {\em maximum} values of the resonance parameter, $q$,
in single-field models of inflation are restricted to the
narrow-resonance regime, $q \leq 1$.

The restriction to the narrow-resonance regime for single-field models of
inflation may be contrasted with the situation in multi-field models.
Consider, for example, the simple model
\beq
V (\phi, \chi) = {\ts{1\over2}}{m^2} \phi^2 + {\ts{1\over2}}g  \phi^2
\chi^2 .
\label{Vpc}
\eeq
In this case, the resonance parameter for the $\chi$ field becomes
\beq
q = \frac{g \varphi_0^2}{4 m^2} \sim g \times 10^{10},
\label{qchaotic}
\eeq
where the last expression on the right holds for the chaotic inflation
case.  In this case, CMB observations only
restrict the coupling $g$ through radiative loop corrections, which
suggest $\vert g  \vert \leq
10^{-6}$.\footnote{This constraint does not exist in (softly-broken)
supersymmetric theories. Hence the inflaton  may be
strongly  coupled to other fields and the resonance parameters may be
arbitrarily large \cite{BT98}.} Thus even weak couplings between the
inflaton and the distinct scalar field $\chi$ will produce resonance
parameters well into the broad-resonance regime, with $\vert q \vert
\gg 1$.  {\em Multi-field models of
preheating generically lie within the broad-resonance regime, whereas
single-field models are necessarily limited to the narrow-resonance
regime.}  This difference will be crucial when we consider the
amplification of metric perturbations at preheating later.

To finish the discussion of preheating in Minkowski spacetime, we
examine the case of non-periodic inflaton evolution. Perhaps the
simplest and most
unified way to do this is in terms of the $1-D$
Schr\"odinger operator, ${\cal L}$, which is dual to the Klein-Gordon
operator in
Eq. (\ref{eom1}) under the interchange of time and space
\cite{bass98}. Under this
exchange, the unstable modes $k$ for which $\mu_k > 0$ are identified
with the {\em complement} of a subset of the spectrum of ${\cal
L}$ (technically the complement of the absolutely
continuous part of the spectrum, $\sigma_{AC}$ \cite{bass98}). From
this we immediately see why $\mu_k$ always had a band structure in the
cases considered above, since in the dual picture we simply find the
conduction bands of condensed matter models of metals with periodic
potentials --  the Brillouin zones. 

In the case where the inflaton evolution is quasi-periodic, the above
duality enables one to show that the band structure of
the periodic case is destroyed, leaving only a nowhere-dense, {\em
Cantor} set of stable modes \cite{bass98,BT98}. Similarly, in the
case that $\phi$ exhibits partial or complete  stochastic behaviour,
one finds that all modes grow  ($\mu_k > 0$), with measure one
\cite{bass98}. In the dual, condensed matter picture, this simply
corresponds to the famous {\em Anderson localization} of
electron eigenfunctions  in metals with random disorder \cite{and58}. 
This was found independently by Zanchin {\em et al.} \cite{noise},
who also showed that it leads to an increase in $\mu_k$ over the
purely periodic case at the same $q$, i.e. noise enhances preheating.

An important application of these results is the case when  $\phi$
evolves chaotically in time, as is expected when there are many
fields coupled  to the inflaton \cite{CL96}. In the
strongly-coupled limit one can show that the $\phi$ motion becomes 
stochastic and hence one knows that preheating is enhanced in 
this region of the  parameter space \cite{BT98}.

\ssec{Unperturbed FRW background}

Now consider what happens for preheating in an expanding FRW (unperturbed)
universe with scale factor $a(t)$, rather
than in Minkowski spacetime.  In this case, a generic equation of motion
for modes coupled to the inflaton takes the form
\beqn
\nonumber \left[ \frac{d^2}{dt^2} + \Omega_k^2 (t) \right] \left\{ a^{3/2}
(t) \chi_k \right\} = 0 ,&& \\
 \Omega_k^2 (t) = \frac{k^2}{a^2 } + m_\chi^2 + g \varphi_0^2 (t)
f (\omega t)&& \nonumber\\
 - \frac{3}{4} H^2 - \frac{3}{2} \frac{\ddot{a}}{a} + \lambda_\chi
\Sigma (t), &&
\label{eom2}
\eeqn
where $H = \dot{a}/a$ is the Hubble constant. Note that both the
physical momentum, $k/a(t)$, and the amplitude of the oscillating
inflaton, $\varphi_0 (t)$, now redshift with the expansion of the
universe.  Note also that Floquet's
theorem no longer applies, since the terms within $\Omega_k^2 (t)$ are
no longer simply periodic.  This  translates into the physical result
that {\em no resonant amplification occurs in an expanding universe
for models restricted to the narrow-resonance regime.}  In
particular, single-field models of inflation
which include a massive inflaton fail to produce any
highly-efficient, resonant particle production
at the end of inflation, when the expansion of the universe is
included.\footnote{
An exception comes with the model of a {\em massless},
quartically self-coupled inflaton which is conformally coupled to
the curvature, for which weak resonance of the
$\delta \varphi$ fluctuations survives in an
expanding universe, again with $\mu_k \sim q \omega \leq \omega$, due
to the conformal invariance of the theory \cite{KLS2,DBDK}.
}

Explosive particle production is still possible in an expanding universe,
however, for multi-field models in the broad-resonance regime, with
$q \geq 10^3$
\cite{num,KLS2}. (For models in which the $\chi$ field has a mass
comparable with the inflaton's mass,
resonant growth occurs for $q \geq 10^4$.)  In this case, the growth of
the $\chi_k$ modes does
not take a simple exponential form; instead, certain modes $\chi_k$ jump
{\em stochastically} each time the oscillating inflaton passes through
zero, but evolve quasi-adiabatically in between these sudden jumps
\cite{KLS2}.
To understand these jumps, it is convenient to work in terms of an
adiabaticity parameter,
\beq
r \equiv \frac{\vert \dot{\Omega}_k \vert}{\Omega_k^2} .
\label{r}
\eeq
The large amplification of the modes $\chi_k$ occurs during those periods
when $r > 1$.  For definiteness, consider again the model of Eq.
(\ref{Vpc}).  In an
expanding universe, for early times after the start of the inflaton's
oscillations, the inflaton will evolve as
\beq
\phi (t) = \varphi_0 (t) \sin (mt) \approx  \varphi_0 (0)
{\sin (mt)\over mt} ,
\eeq
and an effective resonance parameter for the $\chi_k$ modes may be
written
\beq
q_{\rm eff} (t) = \frac{g \varphi_0^2 (t)}{4 m^2} .
\eeq
Then, for $q_{\rm eff} \gg 1$, we have, to leading order in $t^{-1}$,
\beq
r \approx \frac{q_{\rm eff}^{-1/2}}{2} \frac{\cos (mt)}{\sin^{2}
(mt)}.
\eeq
Given that $q_{\rm eff} \gg 1$, we see that we will only get strongly
non-adiabatic evolution of the modes $\chi_k$ near the zeros of $\phi
(t)$. Analytic estimates exist for $\mu_k$ in this case \cite{KLS2}.
Two points are of special interest. First, $\mu_k$ is largest for
$k = 0$, with $\mu_k \propto \ln(1 + 2\exp[-\pi k^2/(2 a^2 m^2
\sqrt{q_{\rm eff}})] + {\cal P})$.
Hence the amplification of super-Hubble modes is present even
in the absense of metric perturbations.  Second, $\mu_k$ depends on
the phase of the wavefunction of $\chi$, through the term ${\cal
P}$ -- a
purely quantum effect. The change in the phase between the zeros of
$\phi$ is much larger than $2\pi$, and hence the phase (mod $2\pi$)
is a pseudo-random variable of time. This stochastic behavior is
transmitted to $\mu_k$.  

Thus, for the broad-resonance regime, strong amplification is
still possible at preheating, though it is no longer of a simple
parametric resonance kind.  Nevertheless, many concepts from the pure
Floquet theory remain helpful, in particular $q_{\rm eff}$ and
$\mu_{k,\rm eff} (t) \equiv \chi_k^{-1} \dot{\chi}_k$
\cite{num,KLS2,NCI}.  As in the Minkowski spacetime case, for $q_{\rm
eff}
\gg 1$, the effective characteristic exponents may still be large,
$\mu_{k,\rm eff} (t) \geq \omega$, though now they are explicitly
time-dependent.  One way to understand the evolution of the amplified
$\chi_k$ modes in this case is that as a given mode, $k/a(t)$, redshifts
with
the expanding universe, it will pass through a series of broad resonance
bands, $\Delta k$, and get amplified in each one \cite{KLS2,ZHS}. As
in Minkowski spacetime, the widths of these
resonance bands for $q_{\rm eff} \gg 1$ scale as $\Delta k \sim q_{\rm
eff}^{1/4}$ \cite{KLS2}. Similarly, the widths of resonance bands for
models with a strong negative-coupling instability in an expanding
universe scale as $\Delta k \sim \vert q_{\rm eff} \vert^{1/2}$
\cite{NCI}.

In the old theory of preheating, resonant passing of energy back to the
inflaton only occurs at second order through {\em rescattering}
effects. This was first
discovered numerically \cite{num} and then explained analytically
\cite{KLS2} as arising from mode-mode coupling, which acts as a driving
term
that leads to population of higher-momentum (shorter-wavelength) modes of
the inflaton. This provides a way of stemming the build-up of backreaction
effects and allows the total variances of the fields to be larger
than would be
estimated using only the Hartree or $N \rightarrow \infty$ approximations,
which miss all rescattering effects. This is of relevance to the
discussion of the maximum scale for non-thermal symmetry restoration
\cite{ntsr,ntsrdefects}. In Sec. \ref{sec:MPerts} we show that when the
metric
perturbations are included, they induce rescattering {\em even at linear
order} and again this helps to raise the maximum variances.

Multi-field models, themselves inherently more realistic than models
which include only an inflaton field, lie generically within the
broad-resonance regime, with $q_{\rm eff} \gg 1$.  Resonant growth of
these coupled field modes, then, marks the immediate post-inflation era,
as preheating produces large occupation numbers $n_k$ for quanta in
various resonance bands $\Delta k$.  The highly non-thermal distribution
of particles results from a stochastic process, in which modes $\chi_k$
slide through a series of resonance bands and receive a large,
non-adiabatic \lq kick' each time the oscillating inflaton passes through
a turning point.
In the broad-resonance regime, this strongly non-adiabatic,
rapidly-growing behavior of certain modes $\chi_k$ remains robust even
when expansion of the universe is included.

As the remaining discussion will demonstrate, it is not only field modes
$\chi_k$ which enjoy such robust resonant behavior at preheating; metric
perturbations do as well.  Moreover, the amplified metric perturbations in
turn provide one more source to induce increased particle production, by
means of gravitational rescattering.  By approaching the immediate
post-inflation epoch in a fully self-consistent manner, the surprising
behavior of preheating becomes all the more striking.

\section{Conceptual foundations of the metric perturbation resonances}
\l{sec:Concept}

Before we embark on a detailed study of the amplification of
metric perturbations during preheating, it is perhaps
appropriate to highlight
the important conceptual differences between metric
perturbation evolution in the
case of a single field and the case involving many fields interacting
{\em both} via gravity and other forces.\footnote{The case of many
non-interacting fields is well studied, but excludes preheating, and 
is typically restricted to slow-roll. See e.g. \cite{multi}.} 

As we will see below (see Eq. (\ref{multi1})), Einstein's field equations
couple
metric perturbations $\Phi_k$ directly to field fluctuations $\delta
\varphi_{Ik}$, for {\em all} wavenumbers $k$.  This means that metric
perturbations can in general be amplified at any $k$.  On small scales,
$k/a \geq H$, such resonances could lead to nonlinear effects such as
pattern and primordial black hole formation \cite{SP98,PBH}.  On much
larger scales, $k/a \ll H$, there is in addition the question of the {\em
causal} amplification of such long-wavelength modes $\Phi_k$.

The schematic in
Fig. (\ref{fig:foundation}) shows the five main points
which are involved in a conceptual understanding of 
metric perturbation resonances in strong preheating. All except causality
are related to the existence of multiple fields in realistic preheating
models, and are relevant for large-$k$ and small-$k$ resonances alike.

\begin{itemize}

\item \underline{\em Non-gravitational forces may dominate}

In the case of fields interacting through non-gravitational forces, the
coupling constants may naturally be large, implying that the evolution of
metric perturbations is {\em non-gravitationally} dominated
\cite{BT98}. This can destroy the solution $\Phi = const.$ for the
$k = 0$ growing mode.

\item \underline{\em Large resonance parameters}

In the single-field case, the natural dimensionless resonance parameter
$q$ is order unity or smaller since both the coupling and frequency of
oscillations is controlled by the same constant of the theory (e.g. the
effective mass or self-coupling). In the multi-field case, the couplings
between fields
may differ from the inflaton frequency of oscillation $m$ by orders of
magnitude,
and hence the associated resonance parameters $q_I \equiv g_I
\varphi^2/m^2$ can naturally be very large.

\item \underline{\em Entropy perturbations}

In the multi-field case, entropy perturbations are generic (see
discussion above Eq. (\ref{ent})).
In the presence of strong non-gravitational interactions
and the associated explosive transfer of energy between fields
with different effective sound speeds, these entropy perturbations
are resonantly amplified, thereby invalidating the use of standard  
conserved quantities to calculate $\Phi$.

\item \underline{\em Causality is not violated}

Causality is not violated by resonant amplification of the $k \sim 0$
modes. This is discussed further in Sec. \ref{sec:Causal} in terms
of the unequal-time correlation function for the perturbed energy density.
The field equations are  relativistic and hence
causality is built into their solution. The
apparent violation of causality arises essentially from the \lq
initial conditions' at the start of preheating:  the homogeneity and
coherence of the zero-mode of the inflaton, which is required to
solve the horizon problem.

\end{itemize}

\begin{figure}
\epsfxsize=3.5in
\epsffile{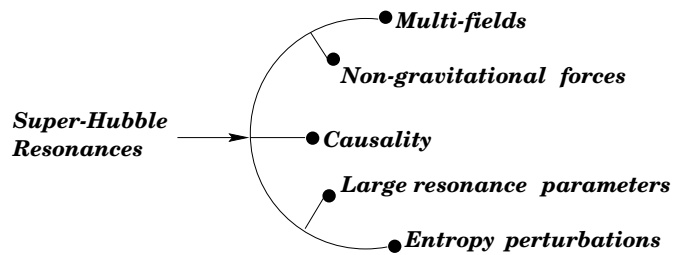}
\vskip 2ex
\caption{Conceptual points of importance in understanding the origin of
the metric perturbation resonances in strong preheating.}
\l{fig:foundation}
\end{figure}

Note that here and throughout this paper, we use the term
\lq\lq super-Hubble" to refer to modes with wavelengths longer than
the Hubble radius, that is, with $k/a < H$.  It is common,
though somewhat misleading, to use the term \lq\lq superhorizon" for this.
As is well-known, the Hubble
radius marks the speed-of-light sphere in any given cosmological
epoch, but can be dramatically different from the actual particle
horizon.  Of course, by solving the horizon problem, inflation
generically produces a particle horizon which is many orders of
magnitude larger than the Hubble radius
during post-inflationary periods in the universe's history.  As
emphasized in this section and below, the \lq\lq super-Hubble"
resonances described in
this paper are thus amplified fully causally; their wavelengths remain
smaller than the true particle horizon after inflation, though they
stretch beyond the Hubble radius at the end of inflation.  Related points
have been made in the study of the backreaction of gravitational waves
\cite{abramogw}:  not only are the super-Hubble modes consistent with
causality and locality, but if one imposes a Hubble-scale infrared cut-off
and excludes these modes, then energy conservation is violated.

Throughout we simply write  $k = 0$ for the
longest-wavelength modes which we consider.  Physically, the
maximum wavelength of perturbations which can ever be amplified
during preheating corresponds to the correlation length of the
coherent inflaton condensate.  Inflation guarantees that the
inflaton background field will be coherent (and thus will oscillate
at preheating with a spacetime-independent phase) over vastly
larger length scales than the Hubble radius at preheating.  This is
what explains the causal amplification of super-Hubble
perturbations.  Thus, where we write $k = 0$, we really intend
something more like $k \sim 10^{-28}$, corresponding to
length-scales well within the Hubble radius during inflation, which
then were stretched coherently during the inflationary epoch.  Of
course, in our numerical simulations, these exponentially-tiny
wavenumbers remain practically indistinguishable from $0$.  We
return to the question of the inflaton condensate's coherence
length below, in Sec. \ref{sec:Back}.

\section{Metric Preheating} \l{sec:MPerts}

With the points of the previous section clearly in view, we may
proceed to study in detail  the behavior of metric perturbations
during preheating. We use the gauge-invariant formalism of \cite{MFB} to
investigate the evolution of scalar perturbations of the metric.
(The amplification of tensor perturbations, or gravitational waves, at
preheating has been considered in \cite{gw}.)
In longitudinal gauge, the perturbed metric takes a particularly simple
form, expressed in terms of gauge-invariant quantities:
\beq
ds^2 = \left( 1 + 2 \Phi \right) d t^2 - a^2(t)\left( 1 - 2
\Psi \right) \delta_{ij} dx^i dx^j \,,
\label{ds2}
\eeq
where $\Phi$ and $\Psi$ are gauge-invariant potentials.
On length scales shorter than the curvature
scale (which itself is longer than the Hubble length), preheating in an
open universe proceeds with little difference from the spatially-flat
case (see the papers by Kaiser in \cite{DBDK}), so we restrict attention
in this paper to the case of a spatially-flat background.

\ssec{The single-field case}

To begin, consider a single-field model, with
the inflaton field separated into a
background field and quantum fluctuations, $\phi (x) = \varphi
(t) + \delta \varphi (x)$,
where $\delta\varphi$ is the gauge-invariant fluctuation in the
longitudinal gauge \cite{MFB}.
The Friedmann and Klein-Gordon equations governing the background
quantities are
\beqn
&& H^2 = {\ts{1\over3}}\kappa^2\left[ {\ts{1\over2}}
\dot{\varphi}^2 + V (\varphi) \right]\,, \label{fried} \\
&&\ddot\varphi + 3H \dot\varphi +
V_{\varphi} = 0 \,,
\label{bkgd1}
\eeqn
where $V_\varphi\equiv\partial V/\partial\varphi$ and
$\kappa^2 \equiv 8\pi G = 8 \pi M_{\rm pl}^{-2}$.
These may be combined to give:
\beq
\dot{H}=
-{\ts{1\over2}}\kappa^2\dot{\varphi}^2 \,.
\label{bkgd2}
\eeq
The energy-momentum tensor
\beq
T^{\mu\nu}=\nabla^\mu\phi\nabla^\nu\phi-g^{\mu\nu}\left[{\ts{1\over2}}
\nabla^\alpha\phi\nabla_\alpha\phi-V(\phi)\right]\,,
\label{roy1}
\eeq
may be written in the form
\beq
T^{\mu\nu}=(\rho+p)u^\mu u^\nu-pg^{\mu\nu}\,,
\label{roy2}
\eeq
where $u_\mu\propto\nabla_\mu\phi$  is the natural four-velocity
defined by the scalar field,
which coincides in the background with the preferred 4-velocity.
The energy density and effective
pressure are
\[
\rho={\textstyle{1\over2}}\dot{\phi}^2+V\,,~
p={\textstyle{1\over2}}\dot{\phi}^2-V\,,
\]
where an overdot denotes the comoving covariant derivative
$u^\mu\nabla_\mu$.
 Thus in the natural frame, $T^{\mu\nu}$
has perfect fluid form, so that the energy flux and anisotropic stress
vanish in this frame: $q_\mu=0=\pi_{\mu\nu}$.
In any other frame which is close to
$u^\mu$, i.e., $\tilde{u}^\mu=u^\mu+v^\mu$, where $v_\mu u^\mu=0$,
we have to linear order
\beq
\tilde{\rho}=\rho\,,~\tilde
p=p\,,~\tilde{q}_\mu=-(\rho+p)v_\mu\,,~\tilde{\pi}_{\mu\nu}=0\,.
\label{roy3}
\eeq
Thus, to linear order, the scalar field has no anisotropic stress
in any
frame that is moving non-relativistically relative to the natural
frame.
Writing $T_{\mu\nu}=\bar{T}_{\mu\nu}+\delta T_{\mu\nu}$,
where $\bar{T}_{\mu\nu}$ is the background energy-momentum tensor
and $\delta T_{\mu\nu}$ is the (gauge-invariant) perturbation,
it follows that
\beq
\delta T^i{}_j \propto \delta^i{}_j \,.
\label{roy4}
\eeq
Then the perturbation of the Einstein equations
$\delta G_{\mu\nu}=\kappa^2\delta T_{\mu\nu}$ gives \cite{MFB}
\beq
\left[\nabla_i\nabla^j-{\ts{1\over3}}\delta_i{}^j\vec{\nabla}^2\right]
(\Phi-\Psi)=0~\mbox{ for }~i\neq j \,,
\label{roy5}
\eeq
so that $\Psi=\Phi$.

The gauge-invariant perturbations for this single-field case then satisfy
the equations
\beqn
&&3H \dot{\Phi} + \left( 3 H^2-\vec{\nabla}^2 \right) \Phi =
\nonumber\\
&&~~~~-{\ts{1\over2}}\kappa^2 \left[ \dot{\varphi} \left( \delta
\varphi\right)^{\rd} + V_\varphi \delta
\varphi-\dot{\varphi}^2\Phi \right] \,, \label{roy6} \\
&&\left( \delta \varphi \right)^{\rd\rd} + 3H
\left(\delta \varphi\right)^{\rd} + \left[V_{\varphi\varphi}-\vec{\nabla}^2
\right] \delta \varphi =\nonumber\\
&&~~~~4 \dot{\varphi}\dot{\Phi} - 2 V_\varphi \Phi \,,
\l{roy7} \\
&&\dot{\Phi} + H \Phi ={\ts{1\over2}} \kappa^2 \dot{\varphi}\delta
\varphi \,.
\label{roy8}
\eeqn
Note that the metric perturbations enter at the same, linear order
in Eq. (\ref{roy7}) as
the usual term, $V_{\varphi\varphi} \delta \varphi$, and must
be included in this equation to remain consistent.
Eq. (\ref{roy8}) can be integrated directly to give:
\beq
\Phi = \frac{\kappa^2}{2a}\int a \dot{\varphi}\delta \varphi\, dt\,,
\label{nonmark}
\eeq
which allows us to eliminate $\Phi$ from Eq. (\ref{roy7}) at the expense
of introducing terms which depend on the complete time-history of
$\dot{\varphi}$ and $\delta \varphi$. In the realistic case where the
fields exhibit a  stochastic component to their evolution
\cite{BT98,bass98,noise},
this non-Markovian term in the evolution equation is expected to cause
new conceptual deviations from the simple Markovian equation in
the absence of $\Phi$ \cite{DBDK}.

By Eqs. (\ref{roy1}) and (\ref{roy2}),
the perturbed energy density and pressure in longitudinal gauge are
\beqn
\delta\rho &=&\dot\varphi(\delta\varphi)^{\rd}+V_\varphi\delta\varphi
-\dot{\varphi}^2\Phi \,, \label{roy9}\\
\delta p &=&\dot\varphi(\delta\varphi)^{\rd}-V_\varphi\delta\varphi
-\dot{\varphi}^2\Phi \,.
\label{roy10}
\eeqn
Eq. (\ref{roy9}) shows how the energy density carried by
gravitational fluctuations,
$\delta\rho[\Phi]=-\dot{\varphi}^2\Phi$,
is {\em negative} when $\Phi>0$ \cite{BKM1}.
The entropy perturbation $\Gamma$
is defined by \cite{KS}
\[
p\Gamma\equiv\delta p-\left({\dot
p\over\dot\rho}\right)\delta\rho\,.
\]
Using Eqs. (\ref{roy6})--(\ref{roy10}), we can show that
\beq
p\Gamma= -\left({4V_\varphi
\over3\kappa^2a^2H\dot{\varphi}}\right)\vec{\nabla}^2\Phi\,.
\label{roy11}
\eeq
Thus on all scales $k > 0$, {\em the single scalar field generically
induces
nonzero entropy perturbations}. These are easily
missed if one neglects the
metric perturbations $\Phi$. Slow-roll inflationary conditions will
keep these entropy perturbations negligible during inflation, but they will
grow in concert with the metric perturbations $\Phi$ during the
oscillatory, preheating phase.

An important quantity is the Bardeen parameter,
i.e., the spatial curvature on uniform density surfaces \cite{MFB}
\beq
\zeta = \Phi - \frac{H}{\dot{H}} \left[ \dot{\Phi} + H\Phi
\right]\,,
\label{zeta}
\eeq
which evolves as
\beq
\dot{\zeta} = \left(\frac{2H}{\kappa^2\dot{\varphi}^2}\right)
 \vec{\nabla}^2 \Phi
\label{dotzeta}
\eeq
in the single-field case.
It follows from Eqs. (\ref{fried}), (\ref{bkgd2}),
(\ref{roy11}) and (\ref{dotzeta}) that
\[
\dot{\zeta}=\left({a^2\rho p\over 2\dot{V}}\right) \Gamma \,.
\]
Provided that $\dot{\varphi}$ (and hence $\dot{H}$) is always
nonzero, the $k = 0$ mode of $\zeta$ is conserved for
models with only one scalar field,
and the $k = 0$ entropy perturbation is zero.
However, during reheating, $\dot{\varphi}$
periodically passes through
zero, and this can produce entropy perturbations even on
super-Hubble scales, as noted in \cite{NT1,FB}.
These entropy perturbations will not be strong, since at most weak
resonance occurs on super-Hubble scales in single-field models.

\ssec{The multi-field case}

We now generalize the above discussion and
study the behavior of $\Phi$ for models with
$N$ interacting scalar fields $\phi_I$, where $I = 1,\cdots , N$,
taking the inflaton to be $I = 1$.  Each field may be
split into a homogeneous part and gauge-invariant
fluctuations as $\phi_I (t, {\bf x}) = \varphi_I (t) + \delta \varphi_I
(t, {\bf x})$.
The background equations, Eqs. (\ref{fried}), (\ref{bkgd2}), become
\beqn
&& H^2 = {\ts{1\over3}}\kappa^2\left[ {\ts{1\over2}}\sum_I
\dot{\varphi}_I^2 + V (\varphi_1,\cdots,\varphi_N)
\right]\,, \label{fried'} \\
&&\dot{H}=
-{\ts{1\over2}}\kappa^2\sum_I\dot{\varphi}_I^2 \,,
\label{bkgd2'}
\eeqn
The first question to address in the multi-field models is whether
anisotropic stress can arise at linear level.
Each field defines its natural frame via
$u_I^\mu\propto\nabla^\mu\phi_I$, and these
are all close to each other in a perturbed FRW universe,
and all reduce to the preferred frame in the background.
Given a choice $u^\mu$ of global frame that is close
to these frames, we have $u_I^\mu=u^\mu+v_I^\mu$, where the relative
velocities satisfy $v_I^\mu u_\mu=0$. The energy-momentum tensor
in Eq. (\ref{roy1}) becomes
\begin{eqnarray*}
T^{\mu\nu}
&=& \sum_I\nabla^\mu\phi_I\nabla^\nu\phi_I \\
&&{}-g^{\mu\nu}\left[{\ts{1\over2}}\sum_I
\nabla^\alpha\phi_I\nabla_\alpha\phi_I-V(\phi_1,\cdots,\phi_N)\right]\,,
\end{eqnarray*}
and Eq. (\ref{roy2}) generalizes to
\[
T^{\mu\nu}=(\rho+p)u^\mu u^\nu -pg^{\mu\nu}+u^\mu q^\nu+q^\mu
u^\nu\,,
\]
where the total energy density, pressure and energy flux are
\[
\rho={\ts{1\over2}}\sum_I\dot{\phi}_I^2+V\,,~
p={\ts{1\over2}}\sum_I\dot{\phi}_I^2-V\,,~
q^\mu=\sum_I \dot{\phi}_I^2v_I^\mu\,,
\]
and the
anisotropic stress vanishes at linear order. Since these total
quantities obey the same frame-transformation laws as in Eq.
(\ref{roy3}), it follows that the anisotropic stress vanishes to
first order in any
frame close to $u^\mu$. Then, as before, Eqs.
(\ref{roy4}) and (\ref{roy5}) hold, and we have $\Psi=\Phi$.
We emphasize that the vanishing of anisotropic stresses in a
multi-field universe holds only in linearized gravity
(see \cite{MGE} for the nonlinear equations). If
fluctuations grow to nonlinear levels and relative velocities
become relativistic, then anisotropic stresses will emerge and
could significantly affect the dynamics
(compare \cite{BM}).

The nature of entropy perturbations is another example where the
single-field and multi-field cases bifurcate strongly. In the
multi-field case, it is known that {\em entropy perturbations are
generic}, arising from differences in the effective sound speeds, from
relative velocities, and from interactions (see, e.g.,
\cite{KS,dunsby2}). Since reheating is the conversion
of energy from a coherent field to a relativistic, thermalized gas, it is
clear that entropy will be generated.  Even exactly
adiabatic initial conditions at the exit from inflation will not prevent
the production of entropy perturbations during reheating,
since the fields are interacting non-gravitationally.

The entropy perturbation in a multi-component system is defined by
\cite{KS}
\beq
p\Gamma=\sum_I\left(\delta p_I-c_{\rm s}^2\delta\rho_I\right)\,,
\l{ent}
\eeq
where the `total' effective sound speed is given by
\begin{eqnarray*}
&& c_{\rm s}^2={1\over h}\sum_Ih_Ic_I^2\\
&&~~\mbox{ where }~h_I=\rho_I+p_I\,,~h=\sum_Ih_I\,.
\end{eqnarray*}
Then $\Gamma$ can be related to $\dot\zeta$.
In the two-field case, $\dot\zeta$ is \cite{multi}
\beq
\dot{\zeta} = - \frac{H}{\dot{H}}\vec{\nabla}^2 \Phi
+{1\over2}H\left[{\dot{\vp}_1^2-\dot{\vp}_2^2\over\dot{\vp}_1^2+
\dot{\vp}_2^2}\right]\left({\delta\vp_1\over\dot{\vp}_1}-{\delta
\vp_2\over \dot{\vp}_2}\right)\,,
\label{roy12}
\eeq
which shows explicitly how $\zeta$ can grow even for $k=0$.
As we have discussed in Sec. \ref{sec:Preh}, preheating
with more than one scalar field is a period of intensely non-adiabatic
evolution, and we may expect that $\dot{\zeta}$ will deviate strongly
from $0$ even in the $k \rightarrow 0$ limit.  This is precisely the
kind of non-adiabatic growth which we observe below.
This invalidates the use of $\zeta$ as a way of finding $\Phi$, and
one is forced to explicitly integrate the evolution equations to find
$\Phi$, as we do in Sec. \ref{sec:Num}.  Further discussion of these
points is made in Sec. \ref{sec:Back}.

Note also that $\zeta$ may be written
in terms of the Mukhanov perturbative quantities
\beq
Q_I=\delta\varphi_I+{\dot{\varphi}_I\over H}\Phi\,.
\label{roy12''}
\eeq
In the two-field case \cite{NT2}
\beq
\zeta={H\left(\dot{\varphi}_1Q_1+\dot{\varphi}_2Q_2\right)
\over \dot{\varphi}_1^2+\dot{\varphi}_2^2}\,.
\label{roy12'}
\eeq

The linearized equations of motion for the Fourier modes of $\Phi$
are (see \cite{kofpog},
where we have corrected the first equation)
\beqn
&&3H \dot{\Phi}_k + \left[ (k^2/a^2) + 3 H^2 \right] \Phi_k =
\nonumber\\
&&~~~~-{\ts{1\over2}}\kappa^2 \sum_I \left[ \dot{\varphi}_I \left( \delta
\varphi_{Ik}\right)^{\rd} - \Phi_k \dot{\varphi}_I^2 + V_I \delta
\varphi_{Ik} \right] , \label{multi1a}\\
&&\left( \delta \varphi_{Ik} \right)^{\rd\rd} + 3H
\left(\delta \varphi_{Ik} \right)^{\rd} + (k^2/a^2) \delta
\varphi_{Ik}=\nonumber\\
&&~~~~4\dot{\Phi}_k \dot{\varphi}_I - 2 V_I \Phi_k - \sum_J V_{IJ}
\delta \varphi_{Jk} , \l{coupledkg} \\
&&\dot{\Phi}_k + H \Phi_k = {\ts{1\over2}}\kappa^2 \sum_I \dot{\varphi}_I
\delta\varphi_{Ik}\,.
\label{multi1}
\eeqn
Equation (\ref{multi1}) exhibits a simple generalization of the
integral solution of the single field case, Eq. (\ref{nonmark}).
There are, however, important distinctions to appreciate between the
single-field and multi-field cases.  Note first that when more than
one scalar field
exists, cross-terms arise in the equations of motion for each of the field
fluctuations, $\delta \varphi_I$.  That is, the evolution of $\delta
\varphi_I$ depends, even at linear order, on all the other $\delta
\varphi_J$.  These cross-terms have been neglected in previous studies of
multi-field models of preheating, which ignored metric perturbations
\cite{preh1,num,KLS2}.
They are present here, in general, because the
gravitational potential $\Phi$ couples universally with all matter fields,
that is, with the entire $\delta T_{\mu\nu}$.  Each matter field
fluctuation
$\delta \varphi_I$ \lq talks' with the metric perturbations, $\Phi$, which
in turn \lq talks' with all of the other matter field fluctuations.  Given
the form of Einstein's equations, then, each of these field fluctuations
$\delta \varphi_I$ becomes coupled to each of the others.

Though distinct from nonlinear mode-mode coupling of the sort discussed in
Sec.  \ref{sec:Back}, these cross-terms, entering at linear order, can
play a large role in making ``gravitational rescattering" quite effective
at preheating.  We adopt the term {\em rescattering} for these
linear-order cross-terms, whereas it is usually reserved for discussing
nonlinear effects, because the coupled form of the equations in Eqs.
(\ref{multi1a})--(\ref{multi1})  indicates that modes are amplified, and
quanta are produced, due to the coupling between fluctuations of
different fields. The significance of these linear-order rescattering
terms is confirmed in our numerical simulations, presented below.

As discussed in Sec. \ref{sec:Preh}, further differences remain generic
between
single-field and multi-field models at preheating, and these distinctions
result in dramatic differences for the metric perturbations in each
case.  In the single-field case,
resonances, when they survive the expansion of the universe at all, are
quite weak.  Furthermore, the amplitude of the inflaton's oscillations
decays with monotonic envelope due both to expansion of the
universe and to dissipation from particle production.  The right hand side
of
the constraint equation, Eq. (\ref{roy8}), is therefore either weakly
growing or decaying in time.  Hence $\Phi_k$ in this single-field case is
either weakly growing or decaying in time.

In the multi-field cases, we may expect far greater resonance
parameters, $q_{\rm eff} \gg 1$, resulting in much larger
$\mu_{k,\rm eff}$ than in the single-field case.  That is, in the
multi-field case, the system will in general be dominated by {\em
non-gravitational} forces; the couplings $g_I$ between the various
scalar fields are independent of the gravitational sector. 
Moreover, the homogeneous parts $\varphi_I$, $I > 1$, typically
grow during the resonance phase, with the energy coming from the
inflaton condensate, $\varphi_1$. 

\ssec{Homogeneous parts of the fields}

There is a subtlety, however, in how to define $\varphi_I$
for $I > 1$ \cite{KLS2}. One approach is to define them implicitly
as the homogeneous objects, $\varphi_I$, which satisfy the unperturbed
Klein-Gordon equations: 
\beq
\ddot{\varphi}_I + 3H \dot{\varphi}_I +
  V_{I} = 0 \,,
  \label{bkgd1'}\\
\eeq 
where $V_I\equiv\partial V/\partial\varphi_I$. Our main complaint with
this definition is that it is not operational. It does not tell one how
to find $\varphi_I$ from the full fields $\phi_I(t, {\bf x})$. More
important
for the study of preheating, these equations (which simply express
the conservation of the homogeneous part of the energy density) miss the
gauge-invariant energy contributions from the $k = 0$ mode of $\Phi_k$.
While in standard analyses this is unimportant, in preheating, a large
amount of energy is transferred to this mode, which should be
included as a homogeneous backreaction effect. As we stressed in
different terms earlier,  it is important to
appreciate that because gravity has negative specific heat, the
energy flowing into $\Phi_{k = 0}$ actually {\em aids} the growth of
field-fluctuations $\delta\varphi_{Ik}$.  

To include this effect, we define the homogeneous parts of the fields $I
> 1$ as:
\beq
\varphi_I \equiv \delta \varphi_{I,k = 0}.
\l{0def}
\eeq
Thus the background parts of the coupled fields are the homogeneous $k =
0$ portions of the fluctuation spectra.
Since, as we saw in Sec. \ref{sec:Preh}, it is
the small-$k$ modes which grow the quickest for models with broad
resonance, the $\varphi_I$ might grow at least as quickly as other modes
$\delta \varphi_{Ik}$ for $k \neq 0$ but small.  Then the combination
$\varphi_I \delta \varphi_{Ik}$, which acts as a source for $\Phi_k$, 
grows even more quickly than either term alone.  This, in turn, makes
$\mu_{k,\rm eff}$ for the modes $\Phi_k$ greater than that for for any of
the
matter fields.  We confirm this behavior below, in Sec. \ref{sec:Num}.
In the end this is not crucial for  the robustness of the resonances
however. In Sec. \ref{kgdef} we employ the definition implicit in
Eq. (\ref{bkgd1'}) and show that the resonances remain, if not quite
as strong.

Now with the definition (\ref{0def}), even if all $k > 0$ modes of {\em
each} of the fluctuations
$\delta \varphi_{Ik}$ happen to lie within stability bands at a time $t$,
$\Phi_k$ will still grow if one of the homogeneous fields $\varphi_I$ lies
within a resonance band.  It is a simple combinatorial problem to
see that as the total number of fields, $N$, increases, the
probability that {\em all} fields $\varphi_I$ and $\delta
\varphi_{Ik}$ lie simultaneously within
stability bands, and hence that $\Phi_k$ receives no growing source
terms, decreases quickly.

As an alternative operational definition, one might use the
field variances as a basis for
defining the homogeneous fields $\varphi_I$ ($I > 1$)\footnote{See
e.g. \cite{spinodal} for a recent implementation of this
definition in inflation.}, viz.:  
\beq 
\varphi_I \equiv \sqrt{ \langle
(\delta \varphi_I)^2 \rangle} = \left[ (2\pi)^{-3} \int d^3k \vert
\delta \varphi_{Ik} \vert^2 \right]^{1/2}\,.  
\eeq 
While the growth
rate of $\varphi_I$ with this definition {\em may} be less than
that when using Eq. (\ref{0def}), it is sure to exhibit resonance
since the $k$-space integral must cross all the instability bands.
This is reflected in all numerical studies of preheating which show
the rapid growth of the variances \cite{num}. Hence the resonant
growth will still exist for the metric perturbations $\Phi_k$.  
 
Even using spatial averaging of the fields $\phi_I(x,t)$ to define
the background fields $\varphi_I$ would lead to resonant growth of
the $\varphi_I$, since the averaging again covers instability bands
when viewed in Fourier space.\footnote{Actually the situation is
rather more subtle: no known covariant and gauge-invariant
averaging scheme exists, even within {\em classical} general
relativity. However, the above statements are true for 
volume averaging in synchronous coordinates and
will remain true for any reasonable averaging scheme. See  
\cite{MAB} for a discussion of backreaction and
averaging issues relevant to inflation.} 

As we will see in Sec. \ref{sec:Num}, these choices for how to define the
homogenous background fields affect quantitative details, but not the
qualitative behavior, of the resonant amplification of modes $\Phi_k$
during preheating.

In contrast, space-{\em time} averaging, as used
by Hamazaki and Kodama \cite{KH2}, removes the oscillations (and
hence the singularities of their particular evolution equations) of
the fields. This is effectively equivalent to taking the limit $q
\rightarrow 0$ and no resonant production of entropy perturbations
was found. Since the perfect fluid equivalence already exists with
oscillating scalar fields, and non-singular systems exist with which to
study
the problem, we disagree with their procedure and note that
this is perhaps the single most important reason why they did not
find the super-Hubble resonances. Taruya and Nambu \cite{NT2}, on
the other hand, include the inflaton oscillations and find resonant
growth of the entropy perturbations (isocurvature mode).  They
claim however that it has no effect on the adiabatic mode, which
they claim is constant. Further, they do not allow the resonance to
go nonlinear and hence claim that the isocurvature mode decays to
zero, essentially leaving no observable traces. Contrary to this,
our explicit simulations show that $\Phi$ is strongly amplified and
once it becomes nonlinear, which on general grounds occurs at or before
the time at which
the matter-field fluctuations do \cite{BKM1}, the role of the
expansion in damping the amplitude of the modes is not at all
clear.

\section{Causality} \l{sec:Causal}

Keeping the important distinction between $H^{-1}$ and the
particle horizon in mind \cite{ER}, we may now consider the role of
causality in the immediate post-inflationary universe.  {\em
Causality involves
correlations across spatial distances.}  As emphasized in \cite{causal},
this is best studied via unequal-time correlation functions in real
space, rather than in Fourier space.  In particular, causality requires
\beq
\langle \delta T_{00} ({\bf r}, \eta) \delta T_{00} ({\bf 0},
\tilde\eta) \rangle = 0 \>\> {\rm for} \>\> r > \eta + \tilde\eta
\,,
\label{causaleq}
\eeq
where $\eta$ is conformal time ($ad\eta=dt$). For simplicity, we
assume that one of the fields $\phi_J, J > 1$, dominates preheating
($\mu_J > \mu_I,~\forall I$).  The correlator (\ref{causaleq}) is
then essentially controlled by the single field $\phi_J$. 
From Eqs. (\ref{roy8}) and (\ref{roy9}), Eq. (\ref{causaleq}) shows
that it is sufficient to study the unequal-time correlation functions
of the field fluctuations, since each term within $\delta T_{00}$
will be proportional to the
field fluctuations to a good approximation, for modes within a
resonance band.  Thus, the function we need to study is $\Delta ({\bf r},
\eta, \tilde\eta) \equiv \langle \delta \varphi ({\bf r}, \eta) \delta
\varphi ({\bf 0}, \tilde\eta) \rangle$, i.e.,
\beq
\Delta ({\bf r}, \eta,\tilde\eta) = \int \frac{d^3 k}{(2\pi)^3} e^{i {\bf
k} \cdot {\bf r}} \delta \varphi_k^* (\eta) \delta \varphi_k (\tilde\eta
) .
\label{causal2}
\eeq
Inside a resonance band, these modes will grow as $\delta \varphi_k
(\eta) = P_k (\eta) e^{\mu_k \eta}$, where $P_k (\eta)$ is a
quasi-periodic, decaying function of time.  Evaluating Eq.
(\ref{causal2}) in the saddle-point approximation, and using Eq.
(11.4.29) of \cite{Absteg}, yields
\beq
\Delta ({\bf r}, \eta, \tilde\eta) \approx \left[ P_k^*
(\eta) P_k (\tilde\eta) e^{\mu_k ( \eta + \tilde\eta) }
\right]_{k_{\rm max}} \frac{ \exp ( - r^2 / \xi^2 )}{2\pi^{3/2} \xi^3} ,
\label{causal3}
\eeq
where $k_{\rm max}$ is the wavenumber at which $\mu_k$ is maximum, and
\beq
\xi^2 \equiv 4 (\eta + \tilde\eta) \left| \frac{ \partial^2
\mu_k}{\partial k^2} \right|_{k_{\rm max}} .
\label{xi}
\eeq
Causality thus places a constraint on the characteristic exponent:
\beq
\left( \mu_k \left| \frac{\partial^2 \mu_k}{\partial k^2} \right|
\right)_{k_{\rm max}} < \frac{1}{4} .
\label{causal4}
\eeq
Note that causality restricts the {\it shape of the spectrum} of
amplified modes, but not directly the wavelength of the
perturbations. The
approximation sometimes used within studies of
preheating, that the distribution of amplified modes falls as a spike,
$\delta (k - k_{\rm resonance})$, violates causality, since this requires
that the field fluctuation contain correlations on all length scales,
even for $k/a \gg H$.
Thus, subject to Eq. (\ref{causal4}), the amplification of
long-wavelength perturbations may proceed strictly causally.  This
raises the question of whether or not metric perturbations, amplified
at preheating, could affect observable scales even today.

The result in Eq. (\ref{causal4}) is of course limited by the single-field
and saddle-point approximations, and also by the fact that the true growth
in resonance bands will not be so cleanly exponential for multi-field
cases in an expanding universe, since, as discussed in Sec.
\ref{sec:Preh},
realistic models of preheating have time-dependent $\mu_{k,\rm eff} (t)$.
Thus the remaining exponential tail in Eq. (\ref{causal4}) is an artifact
of the various approximations used.
The key point is that, regardless
of dynamical details and approximation methods,  causality is not a
question which can easily be answered in Fourier space, and {\em no}
naive estimates of constraints on smallest $k$ modes which could be
amplified
causally can be made, e.g. those based on comparisons of $k$ with $H$.

The point we wish to emphasize is that the field equations are relativistic,
hence causality is built in.\footnote{We thank John Barrow for making
this point to us.}
The governing equations of motion, by themselves,
make no distinctions between $k/a < H$ and $k/a > H$ regimes.  A solution
which satisfies the initial and boundary conditions must, therefore, be
causal [see Fig. (\ref{fig:causal})].  
No matter or energy is moved superluminally to produce the super-Hubble
resonances at preheating.
Any apparently acausal behaviour
must instead be due to the initial conditions, which, as we have
emphasized above, are crucial for stimulating the super-Hubble 
resonances we have discussed.  The resonant amplification of very
long-wavelength perturbations at preheating stems from the initial
coherence of the inflaton condensate at the end of inflation.  We return
to the question of the inflaton condensate's coherence below, in Sec.
\ref{sec:Back}.

As a final demonstration of this point, let us consider an inflaton
distribution which is not a delta-function at $k = 0$ but is rather
distributed over $k$-space. For clarity, let us ignore the metric
perturbations completely.  In this case, the evolution equation
for $\chi_k$ with a $g\phi^2\chi^2$  coupling is of the form:
\beqn
&&\ddot{\chi}_k + 3H\dot{\chi}_k + \frac{k^2}{a^2}\chi_k
\nonumber\\
&&~+ {g\over(2\pi)^6} \int\!\! \int d^3k'
d^3k'' \phi_{k'} \phi_{k - k'} \chi_{k - k' - k''} = 0 .
\eeqn
We see immediately that if $\phi_k = 0$ for $k < k_{\rm crit}$, then there
{\em is no} resonant growth of $\chi_k$ modes for $k < k_{\rm crit}$
since the inflaton is not correlated on those scales. From this
it is easy to see that $\chi_k$ modes are amplified by resonance only up
to the maximum scale on which the inflaton is correlated. Of course,
inverse cascades (see Sec. \ref{sec:Back}) can in principle amplify
longer-wavelength modes, but this is another, essentially nonlinear,
process.

\begin{figure}
\epsfxsize=3.5in
\epsffile{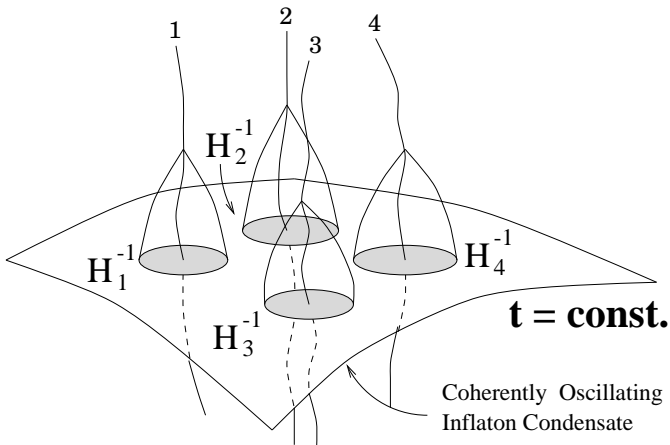}
\caption{Schematic plot illustrating how super-Hubble modes can be
amplified without violating causality. To solve the horizon problem the
inflaton must be correlated on vastly super-Hubble scales at the end
of inflation, and hence this zero mode oscillates with a
spacetime-independent phase. During preheating, fields at each spacetime
event react only to physics in their causal past, which since the end of
inflation may be very small. However, because the inflaton condensate is
exactly the same over vastly larger regions than $H^{-1}$ at the end of
inflation, super-Hubble modes can patch together the inflaton
behavior
in each region to reconstruct (``feel")  the globally oscillating
condensate and hence grow resonantly. Due to the isometries of the
spacelike surfaces of constant $\phi$, all such patches are equivalent
and can be mapped onto one another by a translation.  Only once
second-order backreaction effects destroy the coherence of the inflaton
oscillations are these super-Hubble resonances weakened and perhaps
ended.}
\l{fig:causal}
\end{figure}

The linear case which we solve numerically is qualitatively very
different from the fully nonlinear regime where solutions are not known,
as occurs for example in the studies of topological defects
\cite{defect}. In that case, causality can be imposed as a constraint to
determine whether a proposed ansatz is physical (i.e. on-shell) or not.
This has been used to great effect in recent years in studies of the
evolution of the energy-momentum tensor for causal defects, and in that
context, where perturbations {\em must} come from displacement of
energy, causality sets up strong constraints on the possible metric
perturbation power spectrum.  This is described in detail in
\cite{causal,defect}. 

\section{Numerical results for the 2-field system} \l{sec:Num}

We now study a two-field model involving a massive inflaton, $\phi$, and
a distinct, massless scalar field, $\chi$, with the potential
\beq
V(\phi, \chi) = {\ts{1\over2}}{m^2} \phi^2 + {\ts{1\over2}}{g}\phi^2
\chi^2 .
\label{pot1}
\eeq
Taking $\chi$ to be massless is realistic in the sense that there are
a large number of effectively massless degrees of freedom
($g_* > 100$) at these energies in all post-Standard Model
theories. If $g > 0$, the minimum of the effective potential in Eq.
(\ref{pot1})  is $\chi = 0$, which we assume holds during inflation.
For times after the end of slow-roll in this model, a good
approximation for the evolution of the inflaton background field is:
\beq
\varphi (t) \approx \varphi_0 (t_0){\sin (mt)\over mt}\,.
\label{phiback}
\eeq
This approximation improves with time and is
quite accurate for $\varphi < M_{\rm pl}$. We show it in the inset to
Fig. (\ref{fig:expand}).  This ansatz for the oscillating
background field appears to provide two free parameters:  $\varphi_0
(t_0)$ and $t_0$. Actually, there
exists only one real degree of freedom here, determined by the physical
amplitude of the inflaton at the start of preheating. We have fixed
$\varphi_0 (t_0) = 0.3 M_{\rm pl}$, so
that taking $t_0 \rightarrow 0$ (which gives $\sin
(mt_0) /mt_0\rightarrow1$)
corresponds to a standard chaotic
inflation scenario \cite{num}; we may think of this as a
large-amplitude limit.  
As becomes clear in the following simulations,
the resonances are more effective and modes $\Phi_k$ go nonlinear more
quickly for smaller $t_0$; in particular, see Fig. (\ref{fig:zerotnlq})  
for a simulation with $mt_0 = 1$.
In this case, relevant for simple chaotic inflation, modes $\Phi_k$ go
nonlinear after only $m \Delta t \sim 50$ even for quite weak values of
the coupling $g$ (and hence of the resonance parameter $q \sim O(100)$),
much earlier than field modes $\delta \varphi_{I}$ would go nonlinear in
this simple model when $\Phi$ is neglected (compare with \cite{num}).

Increasing the arbitrary parameter $t_0$ lowers the
initial amplitude of inflaton oscillations, and hence allows us to study
other, low energy-scale inflationary models, some of which we discuss
further in Sec. \ref{sec:Hierarchy}. 
 Parametrizing our equations this way
means that the quoted values for the resonance parameter
\beq
q = \frac{g \varphi_0^2}{m^2}
\eeq
correspond to the values for $t_0 = 0$; larger start-times, $t_0 > 0$,
will correspond to smaller $q_{\rm eff} < q$. Note that here $q$ differs
by a factor $4$ from the $q$ that usually appears in the Mathieu equation.

The Friedmann equation gives an approximate scale for the Hubble
radius at the start of preheating, $H (t_0) \sim 2 \alpha m$, where
$m$ is the inflaton mass, and $\alpha$ gives the physical amplitude
of the inflaton at the start of preheating:  $\varphi (t_0) \equiv
\alpha M_{\rm pl}$.  For chaotic inflation, this yields $H (t_0) / m \sim
0.6$ at the start of preheating; modes with $k/a(t_0) < H (t_0)$ are
thus, roughly speaking, super-Hubble modes.  For our simulations, we
then use the Hubble expansion rate as averaged over a period of the
inflaton's oscillations, $H = 2/(3t)$, which is appropriate for a
massive inflaton.\footnote{ The use of such time-averaging for the
expansion rate and similar background quantities is acceptable in
models dominated by non-gravitational terms, as is the case for the
multi-field model we study here.  This is in general {\em not} a good
approximation when studying gravitational waves or fields with strong
non-minimal couplings to the Ricci curvature scalar, since in these
cases oscillations in the Ricci curvature can lead to strong
resonances \cite{geom}.  } We also take the homogenous portion of the
$\chi$ field to be $\chi = (\delta \chi)_{k=0}$; as discussed above,
using $\sqrt{\langle \delta\chi^2\rangle}$ or the spatial average of
$\chi$ would make quantitative changes, but would not alter the
general features or qualitative behavior presented here. 

In each of the simulations presented below, the initial conditions for
the metric perturbations are taken to be the generic, scale-invariant
spectrum expected to be produced during inflation.  That is, we set
$\Phi_k (t_0) = 10^{-5}$ for all $k$.  As
discussed below in Sec. \ref{sec:Back}, changing these initial conditions
does not, in general, alter the qualitative behavior presented here.
In particular, attempts to fine-tune these initial conditions for $\Phi_k$
so as to avoid nonlinear evolution during preheating are destined to fail.
We have also used the same initial conditions for the matter-field
fluctuations, $\delta \varphi_{Ik} (t_0) = 10^{-5}$ and $(\delta
\varphi_{Ik} (t_0))^\cdot = 0$, where these are measured in units for
which $G = 1$, or $\kappa^2 = 8\pi$.  The initial conditions were chosen
to be scale-invariant (i.e. independent of $k$) to highlight the strongly
$k$-dependent features of the resonance structure.  
We return to the
discussion of the physical impact of different choices of initial
conditions on the resonances below, in Sec. \ref{escape4}.
Note that the figures in the
following subsection plot natural logarithms (base $e$), rather than
base-10.

\ssec{Early Growth of Perturbations}\label{early}

In our numerical simulations, we integrate the coupled, linearized
equations of motion for $\Phi_k$, $\delta \phi_k$, $\chi$, and $\delta
\chi_k$, based on Eqs. (\ref{coupledkg}) and (\ref{multi1}),
 with the specific potential of Eq.
(\ref{pot1}).  The resulting 8-dimensional system of first-order ordinary
differential equations were integrated using a 4th-order Runge-Kutta
routine with variable step-length automatically implemented to ensure
accuracy. Two equations are required each for the evolution of
$\delta \phi_k$, $\delta \chi_k$, and $\chi$, and two for
$\Phi_k$, one with $k > 0$ and one with $k = 0$.
We use Planck units for which $G = 1$, $\kappa^2 = 8\pi$, and a
dimensionless time-variable $z \equiv mt$.  Given the background expansion
$H = 2/(3t)$, the scale factor grows on average
as $a \propto z^{2/3}$.  In other
numerical studies of preheating, in the absence of metric perturbations,
rescaled conformal time $\tau \equiv m\eta$ is often used; in this case, $z
\propto \tau^3$, which facilitates comparisons between our
results and earlier results.

\begin{figure}
\epsfxsize=3.4in
\epsffile{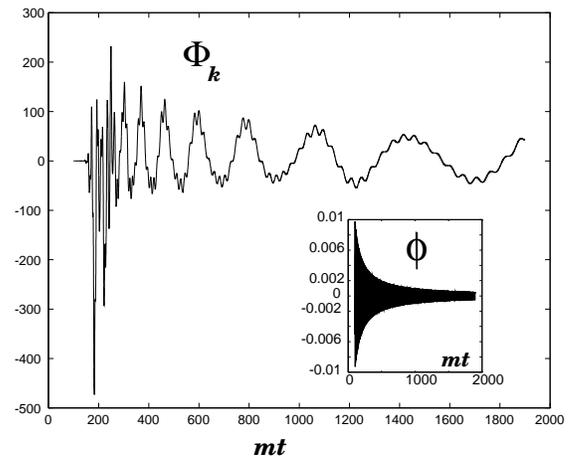}
\caption{$\Phi$ vs $mt$ for $q = 5\times 10^3$, for the $k = 0$
mode. Note the initial resonance which invalidates the linearized
equations of motion within a few oscillations. The linearized solution
then damps away (after $mt \sim 300$) due to the expansion, but this
cannot be taken as indicative of the evolution of the
nonlinear solution. Interestingly there are two characteristic
frequencies of oscillation of $\Phi_k$ after $mt \sim 250$ -- one high
frequency and one low
frequency. The inset shows the evolution of the inflaton
condensate and the damping due to the expansion of the
universe.}
\l{fig:expand}
\end{figure}

Our numerical routine was tested against several limiting cases:
(1) The standard results for preheating when metric perturbations are
neglected, in which the Klein-Gordon
equations for the fluctuations decouple at linear order and the system is
4-dimensional with $\Phi$ set to zero.
In this limit, the standard results of \cite{num,KLS2,NCI} are
reproduced, as discussed further below.
(2) Limiting regions of the parameter space with $q = 0$ were also
checked.  We neglect backreaction on the inflaton and nonlinear
mode-mode rescattering in these simulations, but include the linear-order
coupling which
results between $\Phi$, $\delta \phi_k$ and $\delta \chi_k$.  The truly
nonlinear effects are of course crucial for ending the first, explosive
preheating phase, and we discuss some of these effects below, in Sec.
\ref{sec:Back}.

\begin{figure}
\epsfxsize=3.0in
\epsffile{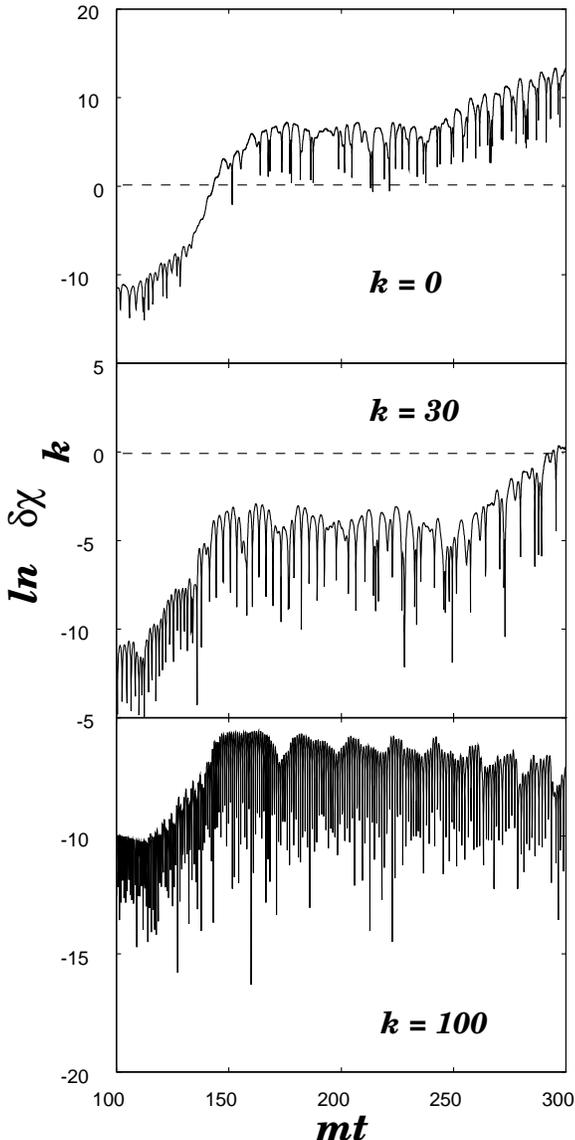}
\caption{The evolution of $\ln \delta\chi_k$ for $q = 8000$ and $k = 0$
(top), $k = 30$ (middle), and $k = 100$ (bottom).  Note the
significant resonant growth of the $k = 0$ mode. }
\l{fig:dchik}
\end{figure}

First consider the growth of metric perturbations $\Phi_k$ in our model.
Fig. (\ref{fig:p2f1}) shows the evolution of two different modes, $k
= 0$ and $k = 20$, as measured in units of $m$. The first is
clearly super-Hubble scaled, whereas the second lies well within the
Hubble radius. As is clear
in this figure, {\em both} modes evolve quasi-exponentially for certain
periods of time; that is, the modes slide in and out of various broad
resonance bands.  The $k = 0$ mode in fact begins to grow considerably
after a time $m \Delta t \approx 40$, after fewer than seven inflaton
oscillations.

\begin{figure}
\epsfxsize=3.0in
\epsffile{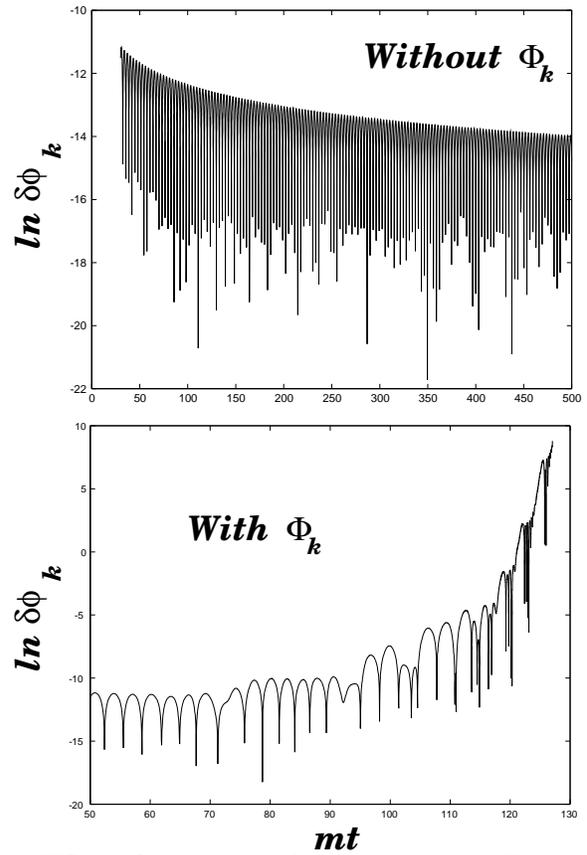}
\caption{$\delta \phi_k$ vs $mt$ evolution, $q = 8000$. Top: note the
complete lack of particle production  in the absence of metric
perturbations. Lower: the particle production due to linear
rescattering via $\Phi_k$ terms is clear. (Note also
that for the inflaton field, we maintain a distinction between $\varphi$,
the oscillating, coherent condensate of zero momentum, and $\delta \phi_{k
= 0}$.  It is the latter which we plot here.) }
\l{fig:withvsno1}
\end{figure}

Figures  (\ref{fig:p2f1}) and (\ref{fig:tnlk}) reveal the strongly
scale-dependent spectrum
for $\Phi_k$ which results from the early, linear-regime evolution of
the coupled system. This is the analog for $\Phi_k$ of the strongly
non-thermal distribution of particles produced in the first stages of
preheating.  We will return to the question of the power spectrum for
$\Phi_k$ during the
preheating epoch in Sec. \ref{sec:Back}.
Of course, the evolution for both modes plotted here beyond $\Phi_k (t)
\sim 1$ cannot be trusted based on our linearized equations of motion, and
nonlinearities will end the resonant growth.  We plot the strong growth
here as an indication that Floquet theory remains an effective tool for
describing the linear regime.  Rather than the simple Minkowski spacetime
results, the effective Floquet indices $\mu_{k,\rm eff} (t)$ become
time-dependent:  the modes $\Phi_k$ slide into and out of resonance bands,
each characterized by a certain exponential-amplification rate.

The coupled scalar field fluctuations $\delta \chi_k$ also experience
dramatic quasi-exponential amplification when the coupling to $\Phi$ is
taken into account, as shown in Figs.
(\ref{fig:dchik},\ref{fig:withvsno2}).

We may now compare the behavior of the matter-field fluctuations, $\delta
\phi_k$ and $\delta \chi_k$, when we include the coupled metric
perturbations, and when we neglect these perturbations as in the standard
preheating literature.  Fig. (\ref{fig:withvsno1}) contains a plot of
the inflaton fluctuations $\delta \phi_k$ with and without including
the coupling to $\Phi_k$, and Fig.  (\ref{fig:withvsno2}) contains a
similar plot for $\delta \chi_k$.

\begin{figure}
\epsfxsize=3.0in
\epsffile{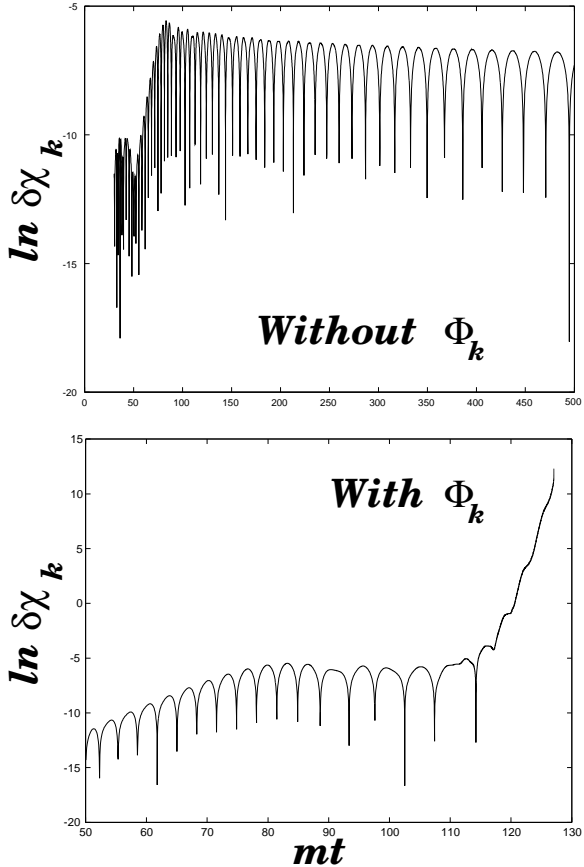}
\caption{Here we compare the evolution of $\delta \chi_k$ vs $mt$ ($k =
0$, $q = 8000$, $mt_0 =
30$) with and without the full metric coupling.
}
\l{fig:withvsno2}
\end{figure}

Note that for the early times plotted here, the behavior of each of these
fluctuations when we neglect $\Phi_k$ matches well the behavior found in
\cite{num,KLS2}. To facilitate comparison, recall that $z
\propto \tau^3$, and that we plot here $\chi = (\delta \chi)_{k=0}$ rather
than $\langle \delta \chi^2 \rangle$.  From Fig. (\ref{fig:dchik}), it
is clear that the $k = 0$ mode of $\delta \chi_k$ dominates the
growth, and hence contributes most to $\langle \delta \chi^2
\rangle$, so this makes for a fair comparison.  The gradual decay of
both $\delta \phi_k$ and $\delta \chi_k$
shown here for early times, when $\Phi$ is neglected, matches the
behavior found in \cite{num,KLS2}.  In these
earlier studies, significant amplification of each of these fluctuation
fields occurred only for $\tau \geq 10$, or, roughly, $z \geq 10^3$.
Instead, when the coupling to the metric perturbations is included, we
find significant growth in each of the field fluctuations well before
these late times.

The strong, resonant growth of both $\delta \phi_k$ and $\delta \chi_k$
confirms that {\em the metric perturbations serve as both a source
and a pump for the matter-field fluctuations.}  We conjectured this above 
and in Paper I, based both
on the form of the perturbed energy density and on the linear-order
gravitational rescattering.  This is a significant new effect which simply
cannot be ignored when studying realistic models of preheating, which
involve more than one scalar field.  Broad-resonance amplification
persists in an expanding universe when $q_{\rm eff} \gg 1$, and this
amplification now affects $\Phi_k$; the fast-growing $\Phi_k$ modes then
further stimulate growth in $\delta \phi_k$ and $\delta
\chi_k$.

Moreover, note that the rate of growth $\mu_{k,\rm eff}$ for $\Phi_{k
= 0}$ is {\em greater} than that for both $\delta \phi_k$ and for $\delta
\chi_k$.  This, too, is consistent with our analysis above, based on the
form of the constraint equation, Eq. (\ref{multi1}), for $\Phi_k$.

Again, we emphasize that the point of these figures is not to study
quantitative details of the field and metric perturbations at late
times, which is impossible using our linearized evolution equations.
Rather it is that quasi-exponential growth of {\em both} of these types of
perturbations will take place early after the start of preheating, and
long before it takes place when the metric perturbations are neglected.

\ssec{Time to nonlinearity}

Given the rapid, quasi-exponential growth of these
perturbations, the most appropriate physical quantity to study is the
time $t_{\rm nl}$ it takes a mode $k$ to saturate its linear theory bound,
that is, $\vert \Phi_k (t_{\rm nl}) \vert = 1$.  Clearly in a system
dominated by gravitation alone, $t_{\rm nl} \sim  H^{-1}_0$, the
current Hubble time, which is much
longer than reheating can last. At preheating, however, in
multi-field models in the broad-resonance regime, we expect instead that
\beq
t_{\rm nl} \sim \mu_{k,\rm eff}^{-1} .
\eeq
Modes lying within a stable band have $\mu_k = 0$ and $t_{\rm nl} =
\infty$, whereas modes within a resonance band will saturate their
linear-theory limit at some finite time.

In Fig. (\ref{fig:tnl1}) we show $t_{\rm nl}$ versus the resonance
parameter $q$ for moderate values of
$q$ and with a cut-off in the integration time of $m \Delta t = 250$.
There is a clear phase transition at $q \sim 2800$, where modes start
going nonlinear for the first time. This is followed by a succession
of narrow stability bands and a second phase transition around $q_*
\approx 6000$ where no more stable modes are found. For $q > q_*$,
essentially all modes go nonlinear
by $m \Delta t \sim 100$, i.e. after only a few inflaton oscillations.

\begin{figure}
\epsfxsize=3.4in
\epsffile{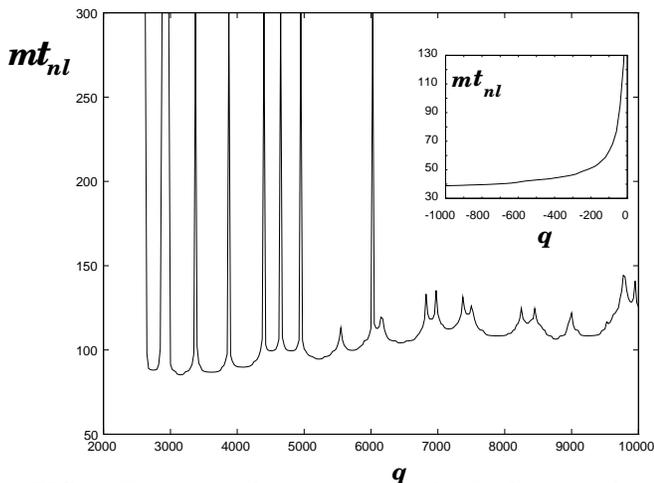}
\caption{Time to nonlinearity $t_{\rm nl}$ vs $q$ for the $k = 0$
mode of $\Phi_k$. Note the
transitions at $q \sim 2800$ and $q_*
\sim 6000$. The inset shows $t_{\rm nl}$ vs $q$ for {\em negative} $q$
and $k = 0$. There are no stable bands and the time to nonlinearity is
short. The numerical solution for the negative-$q$ case is closely
approximated by the simple scaling law
$t_{\rm nl} \sim 1/\sqrt{|q|}$ predicted by Eq.  (\ref{qe:nci}).}
\l{fig:tnl1}
\end{figure}

The inset figure shows $t_{\rm nl}$ for the case of a
negative-coupling
instability amongst the field modes, which produces a more efficient
resonant amplification of all of the perturbations.  Note in particular
that $t_{\rm nl}$ matches very well the prediction from earlier studies of
negative-coupling instabilities that \cite{NCI}
\beq
t_{\rm nl} \sim \mu_{k,\rm eff}^{-1}
\propto \vert q\vert^{-1/2}\,.
\label{qe:nci}
\eeq

In Figs. (\ref{fig:tnl10}-\ref{fig:tnl30}) we show the effect of
decreasing the starting value of $t_0$.  Since
decreasing $t_0$ {\em increases} the effective inflaton oscillation
amplitude, we expect that it should increase the power of the resonance
and hence {\em decrease} $t_{\rm nl}$ and the number of modes lying in
stable bands. This is clearly seen in the figures. 
Note in particular that for most values of $q$ in the range $1500 \leq q
\leq 5000$, and with $5 \leq mt_0 \leq 30$, modes $\Phi_k$ go nonlinear
after $m \Delta t \sim 40 - 100$, well {\em before} any such
nonlinearities appear in studies of preheating which neglect the coupling
to $\Phi$.

Note further that as $t_0$ is
increased, decreasing the strength of the resonance, the resonance band
structure becomes more and more clear.
This behavior of $t_{\rm nl}$ reveals important qualitative agreement with
earlier numerical studies of preheating.  For preheating into massless
fields $\chi$, significant amplification in an expanding universe had only
been found for $q \geq 10^3$ \cite{num}.  The rapid growth of $\Phi_k$
modes found here for our model of the coupled, massless \lq fields' $\Phi$
and $\chi$, falls in this range for $q$.  Moreover, as $q_{\rm
eff}$ increases, the dependence of $t_{\rm nl}$ on $q$ becomes
more fully stochastic, as seen especially in Fig. (\ref{fig:tnl2}).
In this figure, no stable bands persist for $q \geq 3000$, and
$t_{\rm nl}$ varies {\em non-monotonically} with
$q$.  These features again match the qualitative behavior found in
\cite{KLS2} for the stochastic, broad-resonance regime in multi-field
models of preheating that neglect metric perturbations.
We find this same kind of behavior here for the
evolution of $\Phi_k$ with $q \gg 1$.

\begin{figure}
\epsfxsize=3.1in
\epsffile{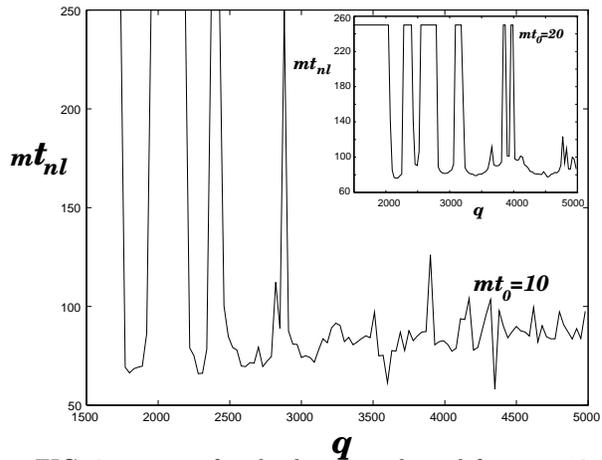}
\caption{$t_{\rm nl}$ vs $q$ for the $k = 0$
mode and for $t_0 = 10$ (main
figure) and $t_0 = 20$ (inset top right). From this it is clear that
varying $t_0$ is not exactly the same as simply varying $q$ essentially due
to the $\dot{\phi}$ terms which are large when $t_0$ is small (the region
where the amplitude of oscillations drops very rapidly). The large $q$,
large $t_0$ regime is different from the smaller $q$, smaller $t_0$ regime
which is closer in spirit to the stochastic resonance regime. This is
evident between this figure and figure (\ref{fig:tnl30}) which shows
ordered resonance and stable bands with $m t_0 = 30$.}
\l{fig:tnl10}
\end{figure}

\begin{figure}
\epsfxsize=3.1in
\epsffile{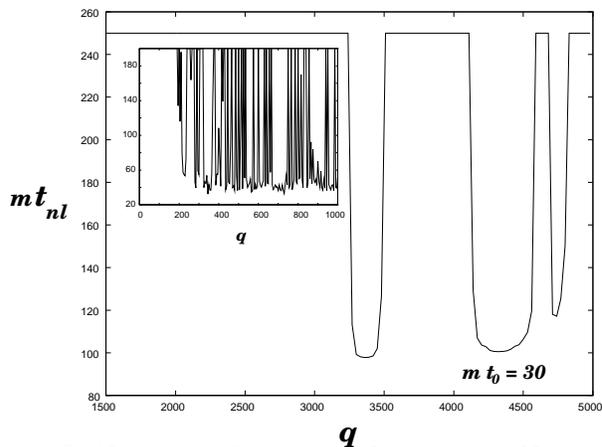}
\caption{$t_{\rm nl}$ vs $q$ for $\Phi_k$, $k = 0$ and $mt_0 = 30$.
Note the ordered resonance bands.  Inset: $t_{nl}$ vs $q$ for $mt_0 =
5$. At this small value of $mt_0$ modes go nonlinear rapidly and no
resonance band structure is apparent.}
\l{fig:tnl30}
\end{figure}

A similar relationship governs the dependence of $t_{\rm nl}$ on $k$, as
shown in Fig. (\ref{fig:tnlk}).  Again, note the absence of stable
bands in $k$-space for
this $q \gg 1$ regime.  Note, too, that the most rapid amplification, and
hence the smallest $t_{\rm nl}$, occur in the $k \rightarrow 0$ limit.
Moreover, this rapid amplification persists smoothly for modes greater
than and less than the Hubble scale, with $t_{\rm nl}$ varying little in the
range $0 \leq k \leq 15$.  This provides a further indication that
$H^{-1}$ is not a fundamental quantity for preheating with more than one
matter field.

\begin{figure}
\epsfxsize=3.1in
\epsffile{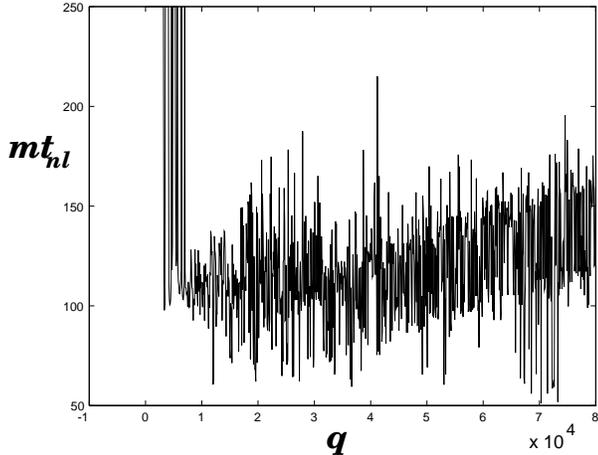}
\caption{$t_{\rm nl}$ over a wide range of $q
> 0$ with $mt_0 = 30$. The essentially stochastic variation of
$t_{\rm nl}$ is clear,
as is the transition to nonlinear behaviour for $q \sim 3000$.
$t_{nl}$ for $q < 0$ is not included here.}
\l{fig:tnl2}
\end{figure}

\begin{figure}
\epsfxsize=3.1in
\epsffile{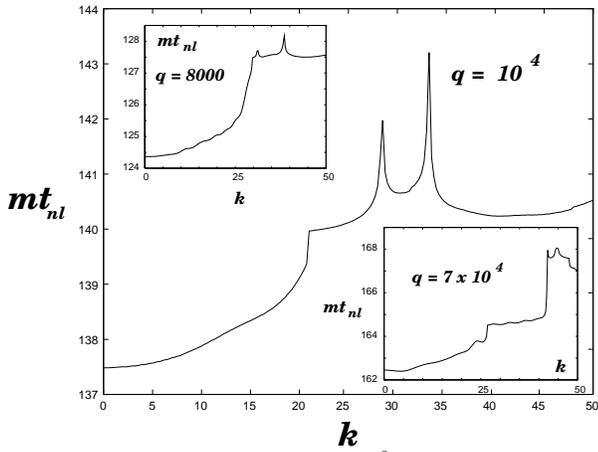}
\caption{$t_{\rm nl}$ vs $k$ for $q = 8\times 10^3$ (top left inset), $q =
10^4$ (main figure) and $q = 7\times 10^4$
(bottom right inset). It is clear that the small $k$ modes are, on
average, the most strongly amplified (with no violation of
causality). This is due to
their low momentum and hence ease of violating the adiabatic condition,
Eq. (\ref{r}). Note that, as expected on general grounds from Floquet
theory, there exist $(q,k)$-parameter regions where  increasing $q$
leads to an increase in $t_{\rm nl}$ showing that the resonance is
not monotonically increasing in strength  with $q$.}
\l{fig:tnlk}
\end{figure}

\ssec{Numerical results with an alternative definition of
the background
$\chi$}\l{kgdef}

A reasonable concern the reader might have is whether the strong
super-Hubble resonances we found in the earlier section are an
artifact of
our definition of the homogeneous component of $\chi$ as the $k = 0$ mode
of $\delta \chi_k$. As already explained, we made that choice so as to be
sensitive to the considerable energy in $\Phi_{k = 0}$, and as such may be
regarded informally as an attempt to include some  backreaction effects.

The standard approach is to require that the background fields satisfy the
unperturbed Klein-Gordon equation, which  simply expresses energy
conservation at zero order. In that case, and for the potential in
Eq. (\ref{pot1}), we
have:
\beq
\ddot{\chi} + 3H\dot{\chi} + g \phi^2 \chi = 0 ,
\l{chizero}
\eeq
which misses the driving terms involving  the homogeneous $\Phi_{k = 0}$
mode.   The $k=0$ mode is not in itself gauge-invariant \cite{bardeen},
but is physical, since, in the longitudinal gauge which we use, all
gauge freedoms are fixed \cite{MFB}.  From a physical point of view, the 
background quantities would correspond to the coherent condensate, with
$k \sim 10^{-28}$ (see discussion at the end of Sec. \ref{sec:Concept}),
and
hence are strictly gauge-invariant.  It is
clearly important to demonstrate that the super-Hubble  resonances persist
for the above, standard, definition of $\chi$. In Fig.
(\ref{fig:zerophi}) we show the evolution of $\Phi_k$ and
$\delta\phi_k$ for $k = 10^{-6}$
for $q = 8000$, $t_0 = 10$. In Fig. (\ref{fig:zerochi}) we show the
evolution of $\delta \chi_k$ and the homogeneous  $\chi$ found from
Eq. (\ref{chizero}) for the same parameter values. Note that although
$\chi$ grows relatively slowly, the resonances in the $k = 10^{-6}$ modes
are still strong, 
with each of the perturbed quantities going nonlinear after $m \Delta t
\sim 65$.

It should be clear that none of the essential predictions have
changed, as
we anticipated earlier in our discussion of the constraint equation.
The $g \phi^2 \chi$ term in Eq. (\ref{chizero}) at large $q$ is
sufficient to ensure
resonant growth of $\chi$, which leads to resonant growth of $\Phi_k$.

In Fig. (\ref{fig:zerotnlq}) we show $t_{\rm nl}$ vs $q$ for $k = 0$
for $mt_0 = 1$,  where $\chi$ satisfies  Eq. (\ref{chizero}). 
Note the fine-structure and the sudden transition to nonlinearity,
characteristic of the earlier $t_{nl}$ vs $q$ figures.
In the usual chaotic inflation scenario, with $m t_0 = 1$, modes $\Phi_k$
go nonlinear after $m \Delta t \leq 50$ even for low-$q$ values of $q \geq
200$, that is, even for very weak couplings $g \geq 10^{-8}$.

\begin{figure}
\epsfxsize=3.1in
\epsffile{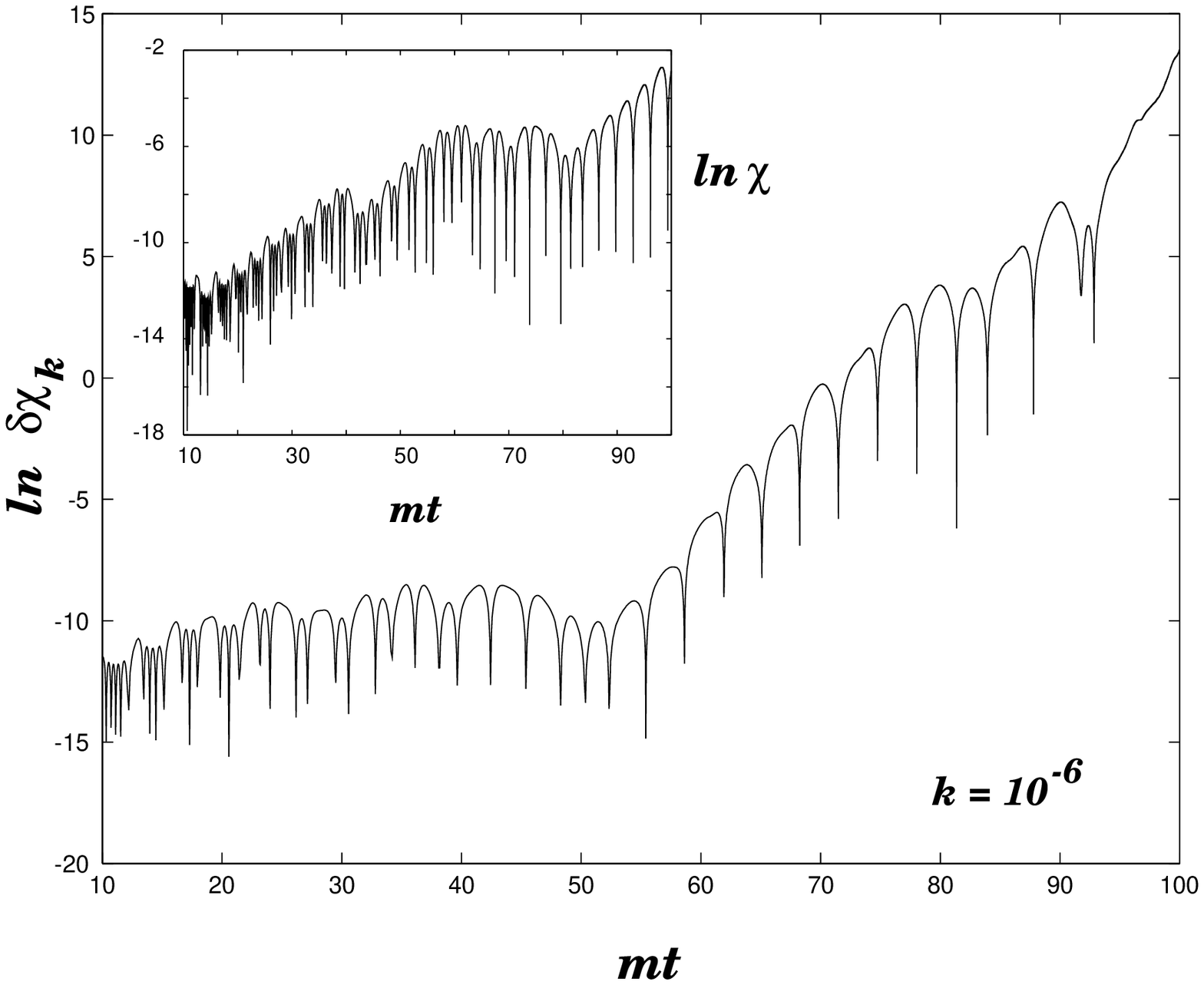}
\caption{
$\delta\chi_k$ vs $mt$ for $q = 8000$ and $k
= 10^{-6}$ and $mt_0 = 10$. {\bf Inset:}
The evoluion of the homogeneous mode $\chi$
vs $mt$ for the same parameter values and
using the evolution eq.
(\ref{chizero}). Notice that although $\chi$
remains small throughout the
simulation, all the perturbations
$\delta\chi_k$, $\delta\phi_k$ and $\Phi_k$
have still gone nonlinear.
This demonstrates that the super-Hubble
resonances are robust
predictions and not sensitive to the
definition of the homogeneous
fields, though the precise times to
nonlinearity are. } 
\l{fig:zerochi}
\end{figure}  

In conclusion, the existence of resonances for $\Phi_k$ is not sensitive
to the choice of homogeneous mode for $\chi$. Saturation of linear
theory and the need for
nonlinear study appears to be a robust prediction of strong preheating,
independent of the definition of the background fields, as long as they
are sensitive to the non-gravitational forces implicit in preheating, as
indeed they must be.

\begin{figure}
\epsfxsize=3.1in
\epsffile{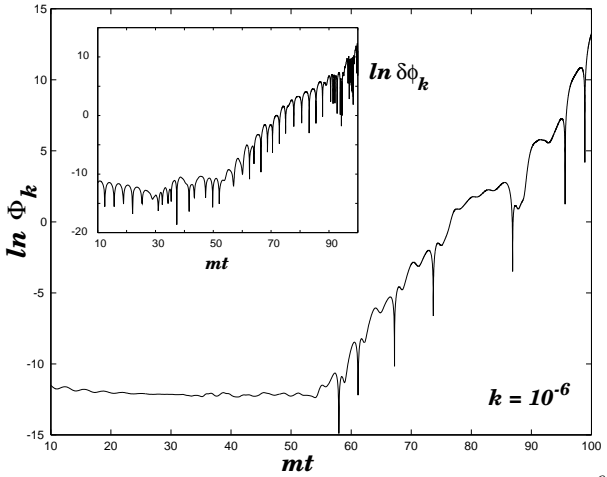}
\caption{
$\Phi_k$ vs $mt$ for $q = 8000$ and $k =
10^{-6}$ and $mt_0 = 10$. {\bf Inset:} The
evoluion of $\delta\phi_k$ vs $mt$ for
the same parameter values.
}
\l{fig:zerophi} 
\end{figure}
\begin{figure}  
\epsfxsize=3.1in
\epsffile{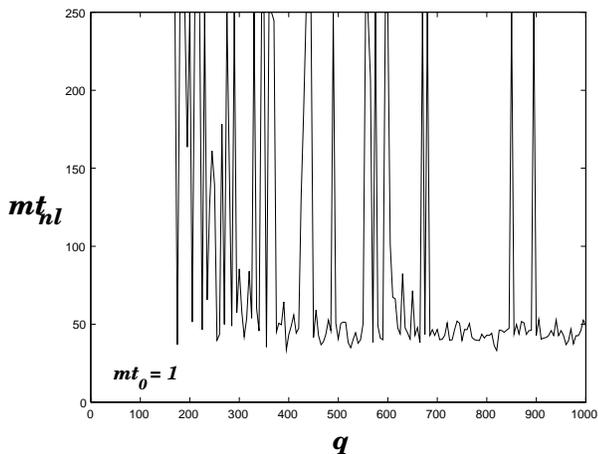}
\caption{$t_{\rm nl}$ vs $q$ for $m t_0 = 1$
and $k = 0$ using the
conservative definition of $\chi$ which
satisfies Eq. (\ref{chizero}). The complex
band structure is still visible.}
\l{fig:zerotnlq}
\end{figure}  

\section{Backreaction, nonlinear aspects and implications}\l{sec:Back}

\ssec{Nonlinearity vs fine-tuning}\l{sec:compare}

In the standard theory based on conserved quantities such as $\zeta$, the
values of $\Phi$ during and after inflation are constrained by a
simple relation such as \cite{breview}:
\beq
\frac{\Phi(t_i)}{1 + w_i} = \frac{\Phi(t_f)}{1 + w_f}
\l{oldcons}
\eeq
where $w \equiv p/\rho$. Clearly, $\Phi(t_f) \gg \Phi(t_i)$ if $w_i
\sim -1$, as occurs in slow-roll inflation. By fine-tuning
$\dot{\phi}$ towards zero during inflation, the amplification of $\Phi$
can be made arbitrarily large. In the standard theory, this doesn't
imply that perturbations have gone nonlinear. Rather, the problem is
run backwards: the $\Phi \sim 10^{-5}$ required by the CMB is used to
determine the value of $\Phi(t_i)$ during
inflation, which therefore puts constraints on the value of $\phi$ during
inflation. No matter what the amount of amplification, the perturbations
can be kept linear by sufficient fine-tuning of the intitial value,
$\Phi(t_i)$.

Have the previous sections simply been a discovery of  a more elegant,
realistic, and refined version of this amplification?  And why are we
claiming that the same fine-tuning cannot be applied in the realistic case
of preheating?  That is, despite the extra amplification due to the
entropy perturbations (which invalidates the use of conserved
quantities such as $\zeta$), why can't all the resonant amplification
be swept under the carpet, and $\Phi$ kept linear, by extra
fine-tuning?

Clearly this is a crucial point.  The extra amplification due to the
quasi-exponential growth of entropy perturbations means that
additional fine-tuning would be needed. This would exacerbate the
already unpleasant fine-tuning of the couplings in many inflationary
models, particularly simple chaotic inflation models.

Yet fine-tuning, while not desirable, is however an issue of taste.  Much
more serious in any attempt to stop the perturbations from going
nonlinear are the following points:

(i) First, fine-tuning the value of $\Phi(t_i)$ downwards during
inflation will typically require making the couplings ($m^2, \lambda$) of
the inflaton field smaller. The dependence of $\Phi(t_i)$ on these
parameters is typically power-law, e.g. $\Delta T/T \propto m/M_{\rm pl}$
in
the quadratic potential model, and $\Delta T/T \propto \lambda^{1/2}$ in
the quartic potential model. However, {\em assuming} (as is true in
simple chaotic inflationary models) that these parameters also control the
oscillation frequency of the inflaton during reheating, we see that the
resonance parameter $q \propto m_{\rm eff}^{-2}$ is significantly {\em
increased} when the effective mass is decreased. This makes the resonance
stronger and {\em increases} the effective Floquet index $\mu_{k,\rm
eff}$.
Since $\Phi_k$ depends exponentially on $\mu_{k,\rm eff}$, it is unlikely
that any fine-tuning  gains made by reducing the coupling constants of
$\phi$ will survive through the end of the resonance in reheating.

(ii) Second, and perhaps of more fundamental importance, concerns the
timing of the end ($t_{\rm end}$) of the initial resonance phase.  Floquet
theory guarantees that the fluctuations will grow resonantly until the
end of this linear phase.
Typically in preheating studies \cite{preh1,KLS2,DBDK}, $t_{\rm end}$ has
been
taken to be the time
when backreaction became \lq important,' which was usually taken on
dimensional grounds to be when $\phi^2 \sim \langle \chi^2\rangle,\>
\langle \delta \phi^2\rangle$, whichever came first.  In Paper I we showed
that when the metric perturbations are included in the analysis, one
expects backreaction to become important in the field
fluctuations and metric perturbations around the same time.  Our
simulations here suggest that at large $q$ (in the multi-field case), it
is often the metric fluctuations $\Phi$ which go nonlinear first.

In the absence of special circumstances (such as a strong $\chi$
self-interaction, $\lambda_{\chi}$), the end of preheating will occur when
second- and higher-order nonlinear processes dominate the linear
resonance, {\em no matter what degree of initial fine-tuning is employed}.
Thus, regardless of special fine-tuning during inflation, preheating in
multi-field models in general will proceed at least until $\Phi$ enters
the nonlinear regime.  For this reason, we consider some implications of
this nonlinearity in the remainder of this section.

\ssec{Mode-mode coupling and turbulence}

Several backreaction effects have been highlighted in the preheating
literature.  Two of them may be approximated with Hartree-Fock or
large-$N$ schemes:  (i) the change to the inflaton amplitude of
oscillation, $\varphi_0$, and (ii) the change to the inflaton
effective mass, $m_{\rm eff} = m^2 + g^2\langle\chi^2\rangle$, and
hence to the frequency of inflaton oscillations
\cite{preh1,KLS2,son}. The first is a strongly non-equilibrium
effect, being governed by non-Markovian dynamics rather than by the
simple Markovian approximation of adding a $\Gamma \dot{\phi}$ term
to the inflaton equation of motion \cite{preh1}. In both of these
cases, the effects derive from the growth of the backreaction term
$\Sigma$ introduced in Sec. \ref{sec:Preh}, $\Sigma_\chi \sim \int
d^3 k \vert \chi_k \vert^2$ and $\Sigma_\phi \sim \int d^3 k \vert
\delta \phi_k \vert^2$.  The implications  of both (i) and (ii) for 
the resonance parameter $q$  appear  straightforward: 
given $q \propto \varphi_0^2/m^2$, the resonance parameter will fall
as the inflaton amplitude falls; it will likewise fall as the
inflaton's effective mass grows. If the effective mass of the
inflaton should decrease, then these two effects would compete. 

As treated within the Hartree-Fock or large-$N$ approximations, these
backreaction effects are essentially independent of $k$; given the form of
$\Sigma_I$, they simply sum up the growth of modes within
certain resonance bands, as a quick way to track the transfer of energy
from the coherent inflaton into various $k$ modes.  As discovered in
lattice simulations \cite{num} and explained analytically \cite{KLS2},
these
approximation schemes thereby miss crucial elements of the nonlinear
evolution.  In particular, $k$-dependent mode-mode coupling can play a
large role before the end of preheating.  It is clear from experience that
any nonlinear field theory may induce mode-mode coupling, and this is the
origin of rescattering in ordinary preheating.

In a Hilbert space, the Fourier transform of a product is simply the
convolution of the individual Fourier transforms; i.e., ${\cal F}(f\cdot
g) = {\cal F}(f) * {\cal F}(g)$.  The conversion of a nonlinear term in
the field equations
into Fourier space therefore yields one or more convolutions of the
spectra of the fields involved. The nonlinear term $g \phi^2 \chi$ in the
equation of motion for $\chi$, for example, becomes the double convolution
in Fourier space (which is symmetric in the arguments of $\phi$ and
$\chi$):
\[
g \int\int d^3{\bf k'} d^3{\bf k''} \phi_{\bf k - k' - k''} \phi_{\bf
k - k'}
\chi_{\bf k} \,.
\]
Assuming statistical isotropy and homogeneity, this reduces
to a double scalar integral with measure $k'^2dk'\cdot k''^2dk''$.  It is
precisely this coupling of modes of different momenta which is missed by
the Hartree-Fock and large-$N$ approximation schemes.

In the case of perturbations around FRW, we must make a splitting of the
scalar matter fields into homogeneous background fields and fluctuations.
When this is done, two kinds of terms arise in performing these
convolutions:  terms for which both $k - k' - k''$ and $k - k'$
are non-zero (i.e. scatterings involving two non-condensate
particles), and terms in which one of the $\phi_k$ particles belongs to
the condensate, with $k = 0$.  The latter type of term is dominant at the
early stages of mode-mode coupling, because the inflaton condensate
occupation number is still high \cite{son,KLS2}.

When metric perturbations are included, similar convolution terms
(and hence rescatterings) appear beyond linear order, now involving
the modes of $\Phi_k$ in addition to $\delta\varphi_{Ik}$.  At first
glance, it might appear simply too difficult to study these terms
analytically, since at second order, the equations describing the
coupled system become tremendously complicated.  The first
complication arises because the variables we use, such as $\Phi$ and
$\delta \varphi$ (and indeed any other known variables), are no
longer gauge invariant.  Thus a gauge-dependent formalism must be
used in practice, although in principle some gauge-invariant
generalizations of these variables might be found \cite{m1}. Further
complications arise from the non-zero pressure and the increased
variety of cross-terms among the many scalar matter fields.  These
issues in gravitational perturbation theory have been explored in a
cosmological setting in studies of dust Einstein-de Sitter spacetime
\cite{m2}. 

These complications, of course, are bound to stymie any analytic progress
on the fully nonlinear evolution of multi-field systems at preheating.
Yet despite these complications, we may make a survey of the {\it kinds} of
terms which will be relevant to the specific question of mode-mode
coupling and rescattering at second order.  At second order, the system
will contain convolutions of two first-order terms, in addition to new
terms formed from background quantities multiplied by explicitly
second-order objects.  Because the background quantities yield
delta-functions in Fourier space, however, these latter terms make no
contributions to mode-mode coupling.  Thus, modulo subtleties regarding
gauge fixing, the nonlinear effects from mode-mode coupling arise from
combinations of terms which are now familiar from the linearized equations
studied above.

Consider, as an example, possible effects of mode-mode coupling for a
single-field model with small-amplitude oscillations ($\varphi_0 \ll
M_{\rm pl}$).  In this case, the linear theory predicts
\cite{NT1} no amplification
of $k \rightarrow 0$ modes of $\Phi$ at preheating.  The constraint
equation at second order for this simple case would take the form:
\beqn
&&\dot{\Phi}^{(2)}_k + H^{(1)} * \Phi^{(1)} + H \Phi^{(2)}_k
= \nonumber\\
&&~~~{\textstyle{1\over2}}\kappa^2
\int{dk'\,k'^2\over(2\pi)^3}\,
\delta\dot{\varphi}^{(1)}_k \delta \varphi^{(1)}_{k-k'} +
{\textstyle{1\over2}}\kappa^2 \dot{\varphi} \delta
\varphi_k^{(2)}\,.
\eeqn
The integral on the right-hand side can transfer power from high-$k$ to
small-$k$ modes (an {\em inverse cascade}), as well as from small-$k$ to
larger $k$ (a forward cascade). Such inverse cascades could in
principle amplify super-Hubble modes of $\Phi$ even in such a
small-amplitude, single-field model, based
on these mode-mode couplings. Inverse cascades of precisely this type are
common in nonlinear systems \cite{invcasc}, and indeed evidence
for them was found recently in a numerical study of preheating
\cite{PE}. 

While still a topic of significant debate in turbulence
circles,  the classical results of Kolmogorov on scaling in
turbulence \cite{kol} suggest  a $k^{-5/3}$ scaling law for the
energy cascade in 3D turbulence. Similar direct cascade behavior has been
found in lattice preheating simulations \cite{num}.  An even more radical
possibility is that of inverse cascades, in which power is transferred
from small to large scales.  Such behavior is believed to occur in energy
transport in 2D hydrodynamics, as well as in particle number in the
nonlinear Schr\"odinger equation, which signals Bose condensation.  In
addition, there is the claim \cite{invcasc} that energy cascades are
inverse in
nature in turbulence and magneto-hydrodynamics with power-law  
spectra $\propto k^{\alpha}$, if $\alpha > -3$, as is the case  
in inflation, although see also \cite{soncasc}. Of course,
in the cosmological context causality must play a crucial r\^ole.
Here, the coherence of the inflaton condensate is irrelevant and
cascades must obey causality, yielding the usual result that the
part of the spectrum added by mode-mode coupling can go on large
scales at best like $k^4$ \cite{causal,BZH}.

Similar terms involving convolutions over $\Phi_k^{(1)}$ and $\delta
\varphi_{Ik}^{(1)}$ would appear at second order in the equations of
motion for both the gravitational and matter-field fluctuations in
multi-field models.  For the matter-field fluctuations, these nonlinear
$\Phi * \delta \varphi_I$ convolutions would therefore add a new source
for rescattering to those considered in earlier studies:  scattering of
$\phi$ and $\chi$ particles directly off the gravitational potential.
This nonlinear mode-mode gravitational rescattering would  complement the
gravitational
rescattering we found already at linear order, in the form of the $V_{IJ}
\delta \varphi_{Jk}^{(1)}$ cross-terms examined in Sec. \ref{sec:MPerts},
ultimately leading to larger variances in the matter-field fluctuations
than the linear theory alone will produce.

\ssec{The post-preheating power spectrum and the CMB}

Perhaps more important, we may expect the results found in earlier
numerical studies of preheating, regarding turbulence and the final shape
of the power-spectrum for fluctuations \cite{num}, to carry over to
the metric perturbations $\Phi$ as well.  As found there for the case of
coupled matter-field fluctuations, turbulence and cascades will
lead, in general, to a re-establishment of  scale-invariance for the
power spectra ($P_I (k)$) for each $\delta \varphi_I$.  In the case
of $\Phi$, then, these mode-mode coupling effects ultimately would
lessen the sharply scale-dependent spectrum which results at the end of
the linear era of preheating, though as discussed in the previous
subsection, causality will limit its effects for very small $k$.
This nonlinear effect on $P_\Phi (k)$
could thus produce a nearly scale-invariant spectrum by the end of
preheating above a threshold value of $k$.  It is crucial to
recognize, however, that this \lq\lq late-time" (end of preheating)
power spectrum $P_\Phi (k)$ would be virtually independent of the
scale-invariant spectrum produced {\em during} inflation \cite{CG97}.
The nearly-flat spectrum measured by COBE, which probes $P_\Phi (k)$
over very small $k$, would therefore be a complicated combination of
these two spectra.

However, based on earlier work on
second-order perturbations and the CMB \cite{DS94}, it is clear that
amplification of super-Hubble modes to the stage where second-order
perturbations cannot be neglected is not compatible in naive
models with the COBE DMR results that $\Delta T/T \sim 10^{-5}$.
Hence some mechanism is required to damp the spectra or limit
the resonances before they can be observationally compatible. 
Some implementations of this requirement are discussed in Sec.
\ref{sec:waysout}. 

An interesting possibility also remains for observational consequences of
$P_\Phi (k)$ as it evolves through its highly nonlinear mode-mode coupling
regime in the later stages of preheating.  This second effect concerns
larger-$k$ (shorter-scale) perturbations.  The turbulent, nonlinear
evolution of $\Phi_k$ modes at preheating could effectively mimic generic
features of the power spectra predicted by topological defect models.  In
particular, the broad-resonance, stochastic amplification to nonlinearity
of certain modes $\Phi_k$ could potentially smear out the
secondary acoustic Doppler peaks in the $\ell > 100$ portion of the
CMB spectrum, if the modes remained nonlinear all the way down to
decoupling. These sharp secondary Doppler peaks are expected in
nearly all inflationary models (when one ignores completely the
behavior of $\Phi$ during preheating), yet are usually absent from
defect models (which feature \lq\lq active," incoherent spectra,
especially in these higher multipoles) \cite{doppler}.

Combined with the spectrum at small $k$, this could produce a
hybrid spectrum for $P_\Phi (k)$: (nearly) scale-invariant at scales
probed by COBE (and hence naively similar to standard inflationary
predictions, though physically of radically different origin), yet
lacking any clear secondary Doppler peaks at shorter scales (and
hence naively similar to standard predictions from defects models,
though again for very different underlying physical reasons).  The
specific effects of the resonant amplification of $\Phi_k$ modes at
preheating on the CMB requires careful study, and is currently under
investigation by the authors.

\ssec{Coherence of the inflaton zero mode}

As we discusssed in Sec. \ref{sec:Causal}, the super-Hubble 
resonances which form the basic element of this paper owe their
existence to the coherent
oscillations of the inflaton zero mode.  These oscillations lead to
stimulated effects that are missing if one treats the inflaton as an
incoherent fluid \cite{KLS2}.  Hence, in addition to the backreaction
mechanisms outlined above, we expect that another important backreaction
event will
be the reduction in the correlation length of the inflaton condensate. This
is expected on very general grounds based on the fluctuation-dissipation
theorem \cite{dennis}:  backreaction will cause the long-range coherence
of the inflaton condensate to be broken.  Eventually the phase of the
inflaton's oscillations will become spacetime
dependent, with its own correlation length that is expected to decrease
rapidly \cite{BBKM}.

As a result we expect the treatment of the inflaton as
an effectively infinite,
coherent condensate to break down on a timescale $t_{\rm coh}$.
However, the reader should note that at first order in perturbation theory
this effect does not appear, and only at second and higher order can the
loss of coherence appear consistently. Hence the super-Hubble 
resonances cannot be eliminated. These higher-order effects, rather,
would decrease the time during which the very-long wavelength metric
perturbations would be amplified, and hence could limit the final
amplitude of these modes.

For times $t > t_{\rm coh}$, as the inflaton correlation length decreases
rapidly and the inflaton begins to lose coherence as higher-momentum modes
get amplified, we expect the resonance for the $k/a \ll H$ modes to weaken
significantly.  This would likely end further resonant growth of the
super-Hubble modes.  From that time on, the only mechanism which
could
amplify the $k = 0$ mode would be inverse cascades.

The coherence issue has been discussed briefly previously.  In
\cite{fuji2} it is argued that the inflaton's coherence is only lost
if the decay rate $\Gamma_\phi$ of the
condensate, considered as a collection of zero-momentum bosons, is
greater than $H$.  In this sense, the expansion of the universe has a
`healing' effect on the coherence. For a coupling of the form in Eq.
(\ref{Vpc}), they argue 
(using only the high-frequency limit to calculate the cross-section
for $\Gamma_{\phi}$) that this ratio is 
\beq
\frac{\Gamma_{\phi}}{H} \approx  1.2 \times 10^5 g^2 ,
\eeq
so that only for $g > 3 \times 10^{-3}$ is coherence lost.  Note that this
does not
depend immediately on $\varphi_0$ and $m$ ($m$ was cancelled in the
cross-section and the average $H$ was used, so $\varphi_0$ did not appear).
If this is correct, then in the simplest models of chaotic inflation, Eq.
(\ref{qchaotic}) predicts that all values of $q \leq O(10^7)$ lead to
coherent inflaton oscillations, regardless of backreaction.  In this
case, the super-Hubble resonances of the $k \approx 0$ modes would
continue until $q$ was significantly reduced via the other
$q$-reducing backreaction mechanisms.  This point was considered
separately in \cite{KK96}, in which it was argued based on
semi-classical quantum gravity that coherence would be lost
rapidly after only a few oscillations, though no detailed arguments were
given.

Note that while the break-up of the coherence of the inflaton
condensate may stunt the super-Hubble resonances, it will not stop
resonances on smaller scales. Further, recent work (see the
second ref. in \cite{noise}) has shown that parametric resonances
persist even in the case where the effective mass is modulated by
spatially inhomogeneous white noise. They further conjecture that
the noise increases the Floquet exponent, $\mu_k$, for all modes. If
this is correct, the break-up of the condensate may not aid in
ending the super-Hubble resonances at all.

Quite apart from these gravitational considerations, yet another
fundamental difference appears relevant between single-field and
multi-field models:  the break-up of the super-Hubble coherence of
the inflaton condensate may proceed more quickly in multi-field
models than in the single-field case.  This stems not only from the
larger couplings between fluctuations in the multi-field case, but
also from the greater number of directions in field space along which
the coupled system could exit inflation.  In other words, the
inflaton condensate coherence length even at the start of preheating
is likely to be shorter in multi-field models than in the
single-field case.\footnote{DK would like to thank Robert
Brandenberger and Richard Easther for helpful discussions on this
point.} Similar effects have been highlighted in recent studies of
tunneling in multi-field models of open inflation, in which the
multidimensional field-space yields large \lq\lq quasi-open" domains
inside the nucleated bubble, rather than a perfectly homogenous
background.  \cite{quasiopen} The coherence of the inflaton
condensate thus remains a crucial quantity in need of further study;
the amplification of super-Hubble modes depends directly upon it
\cite{BBKM}. 

\section{Possible  escape routes  from nonlinearity}
\l{sec:waysout}

Given the severity of the consequences of the super-Hubble resonances
with regard to observational compatibility, primarily with the CMB, a
natural question, and one to which we have already alluded previously in
the paper, is ``can one have large $q$ but still escape the
super-Hubble resonances?''

\ssec{Escape route 1:  secondary phases of inflation}

The no-hair theorem \cite{HE73} assures us that should a second phase of
inflation follow preheating, the large super-Hubble resonances will
be smoothed with time, the universe re-approaching the de Sitter
geometry.

Such secondary phases of inflation are desirable and even expected in
certain models. Consider thermal inflation for example \cite{thermal},
proposed as a solution to the moduli problem with a relatively small
number ($\sim 10$) of $e$-folds. Such a phase of inflation following
preheating would stretch and smooth, but not eliminate, the
signature of the nonlinearity achieved during preheating. The challenge
is to identify the fingerprint that one would expect on the
large-angle CMB in this case.

Secondary phases of inflation are expected in multiple inflationary models
based on supersymmetry and supergravity \cite{subir}. Studies
have already begun examining the effect of the
sudden change in the effective mass of the inflaton in the transition from
one phase to another on the CMB, though until now the change has been
treated neglecting any resonance effects \cite{subir,bsi}.

Finally, it appears that the COBE results prefer a relatively low
energy scale for the production of the anisotropies,
$V^{1/4}/\epsilon^{1/4}  \sim  6.7 \times 10^{16}$ GeV, where
$\epsilon \equiv {1\over2}M_{\rm pl}^2 (V'/V)^2 \ll 1$ is the slow-roll
parameter (see Liddle and Lyth in \cite{infl}).  Such a low energy scale
is a problem if one also
wishes inflation to truly solve the horizon/flatness problem of the
standard cosmology, since this low-energy scale corresponds to a lifetime
several orders of magnitude larger than the Planck time, the ``natural"
time-scale for recollapse \cite{infl}. This is a generic problem of
`low-energy' inflationary models. A first phase of chaotic inflation
which sets up appropriate initial conditions for a second phase of
inflation, which in turn produces the required density fluctuations,
is an alternative \cite{2phase}.

Of course this is not the only solution. The above
objection cannot be made of compactified multi-dimensional models in which
4-dimensional unification occurs around the GUT scale, since then GUT
scale inflation occurs immediately upon exit from the Planck epoch.
In the most extreme version of these models, in which unification occurs
around the TeV scale \cite{gia}, it is, however, not clear at present
whether inflation can be implemented  naturally or not
\cite{mminf,gia2}.

\ssec{Escape route 2: fermionic and instant preheating}

Until now we have considered a model of the post-inflationary universe
based solely on scalar fields. The case of preheating to gauge bosons is
known to be qualitatively similar to the scalar case \cite{gb},
though more complex. Fermionic preheating, on the other hand, is
effectively crippled by the Pauli exclusion
principle, which places the limit $n_k \leq 1/2$ on the occupation number,
as opposed to the $n_k \gg 1$ possible in the scalar case \cite{f1,f2,f3}.

A natural question then is how the metric resonances are affected when
the inflaton can decay {\em only} into fermions at the end of inflation.
Our aim here is to present
evidence that the exclusion principle effectively stiffles the resonant
amplification of metric perturbations at preheating, and thereby possibly
offers a way out from the super-Hubble resonances that occur in the
scalar case.

Consider a massless spin $\ts{1\over2}$ fermion, $\psi$, coupled
to the inflaton via a $h\overline{\psi}\phi\psi$ interaction term.
$\psi$ satisfies the Dirac equation
\beq
[i\gamma^{\mu}\nabla_{\mu} - h\phi(t)]\psi = 0\,,
\l{dirac}
\eeq
where $\gamma^{\mu}$ are the Dirac matrices satisfying the Clifford
algebra relations $\{\gamma^{\mu},\gamma^{\nu}\} = 2g^{\mu \nu}{\bf
1}$. Making
the anzatz $\psi = [i\gamma^{\mu}\nabla_{\mu} + h\phi(t)]X$ (see
\cite{f2,f3}), $X$ satisfies the Klein-Gordon equation with a complex
effective mass [cf. Eqs. (\ref{eom1}) and (\ref{Akq})]:
\beq
\Omega_k^2 = \omega^2_k + qf - i\sqrt{qf}
\l{effmassferm}
\eeq
where $f = \phi^2/\varphi^2_0$ as in Sec. \ref{sec:Preh}, and $q \propto
h^2/m^2$ for a massive but not self-interacting inflaton, while $q \propto
h^2/\lambda$ in the case of a massless,
quartically self-coupled inflaton. The comoving number of created
fermions is \cite{f3}:
\beq
n_k = \frac{1}{2} - \frac{\left[2\omega_k^2\, \mbox{Im}(X_k\dot{X}_k^*) +
\sqrt{qf}\right]}{2\Omega_k}
\l{numferm}
\eeq
with energy density
\beq
\rho_{\psi} = \frac{1}{2\pi^3}\int dk k^2 \Omega_k n_k\,.
\l{rhoferm}
\eeq
Note that most of this energy density is in the form of inhomogeneity
since $n_{k = 0} \leq 1/2$ like all other modes, {\em unless} a
(chiral-symmetry breaking) condensate forms with $\langle
\overline{\psi}\psi\rangle \neq 0$. Assuming that such a condensate has
not formed, one expects  from Eq. (\ref{rhoferm}) that the total
energy radiated into the fermion field will be small,
essentially due to the low occupation numbers forced by the
exclusion principle.

This is borne out by a full closed-time-path analysis of the problem
\cite{f1}, which shows that the ensemble-averaged energy
dissipated to $\psi$ from a massive inflaton is:
\beq
\langle\!\langle\,\rho_{\psi}\,\rangle\!\rangle = \frac{m^2 h^2
\varphi_0^2}{64\pi^2} ,
\l{rhoferm2}
\eeq
which is rather small. This should be compared with the energy density in
scalar $\chi$ particles at $t_{\rm end}$, the end of the linear
resonance, which one can estimate in the Hartree approximation
\cite{KLS2} as $\rho_{\chi}(t_{\rm end}) \sim g |\phi|n_{\chi} \sim
m^2 \varphi_0^2$. This second energy density is much larger than its
fermionic counterpart unless
$h^2 \geq 1$, and this is not allowed by the CMB, unless the theory
is supersymmetric. Further, the energy pumped into the $\chi$ field
predominantly goes into super-Hubble modes, whereas in the fermionic
case, only a small fraction of the total fermionic energy can go into
amplifying super-Hubble modes due to the exclusion principle. This is
exacerbated by the small phase-space volume, $\propto k^3$, at small
$k$. 

The equation of state for $\psi$ must satisfy  $p_{\psi} \leq
\rho_{\psi}$, and hence we are assured that $\delta T^{\mu \nu}$ is small
relative to the energy density in the background. The perturbed
Einstein field equations \cite{MFB}
\beq
\nabla_i(\dot\Phi + H\Phi) \propto \delta T^0_i
\eeq
suggest that it is unlikely that any significant amplification of
super-Hubble metric perturbations exists in
the case of preheating only into fermions.

More interesting is the realistic case where scalar and fermionic fields
coexist, coupled to the inflaton. Examples of fermion-stimulated decay of
the inflaton into bosons exist \cite{f2}, and it is not clear how metric
perturbations evolve in this case.

A particular example is the recent model known as instant preheating
\cite{instant}, in
which the sequence $\phi \rightarrow \chi \rightarrow \psi$ occurs, with
the end point being small numbers of very massive fermions, with masses
$m_{\chi} > 10^{16}$ GeV possible. While it is not our aim to study the
evolution of $\Phi$ in this model, we expect that the resonances discussed
here will not exist in their current form in instant preheating, since the
$\chi$ bosons decay almost immediately (within one oscillation) in the
simplest scenarios, while the resulting fermions are few in number,
and the energy transferred to them is typically rather small.
Still, this energy may grow to dominate the total energy density of the
universe with
time if the inflaton energy redshifts relativistically, $\propto
a^{-4}$. It is difficult to say anything more quantitative, and these
interesting issues are left for future work.

\ssec{Escape route 3: $\chi$ mass and self-interaction}

In our numerical analysis, we limited ourselves to a model of
preheating with a massless scalar field $\chi$. Two $\chi$
characteristics one might expect of $\chi$ are a non-zero mass
($m_{\chi} \neq 0$), and self-interaction ($\lambda_{\chi} \neq 0$).
For the potential
\beq
V(\phi,\chi) = {\ts{1\over2}}m_{\phi}^2\phi^2 +
{\ts{1\over2}}{g}\phi^2\chi^2 +
{\ts{1\over2}}{m_{\chi}^2}\chi^2 + {\ts{1\over4}}{\lambda_{\chi}}\chi^4\,.
\l{eq:fullpot}
\eeq
the perturbed Klein-Gordon equation, Eq. (\ref{coupledkg}),
for $\delta \chi_k$ then becomes
\beqn
&&(\delta\chi_k)^{\rd\rd} + 3H(\delta\chi_k)^{\rd} +
\left[\frac{k^2}{a^2}  + m_{\chi}^2 + g\varphi^2 +
3\lambda_\chi \chi^2\right]\delta\chi_k\nonumber\\
&&~~{}= 4\dot\chi\dot{\Phi}_k
- 2\left[g\varphi^2 \chi + m_{\chi}^2 \chi +
 \lambda_{\chi} \chi^3\right]\Phi_k \nonumber\\
&&~~~~{}-
2g\varphi\chi\delta\varphi_k .
\l{eq:chimass}
\eeqn
From this it is clear that the effective mass of the $\chi$ field depends
on $m_{\chi}$,
$\chi^2$ and $g\varphi^2$ (whether one uses $[(\delta\chi)_{k=0}]^2$ or
$\langle\delta\chi^2\rangle$ for the background field $\chi$ is not very
important at this
qualitative level). In the simple preheating case in which metric
perturbations are ignored, increasing $m_{\chi}$ leads to a sharp
decrease
in the strength of the resonance.
(Effectively it increases the $A$
parameter of the Mathieu equation ($A \propto k^2/a^2 +
m_{\chi}^2/m_{\phi}^2 + 2q$), leading to a rapid decrease in $\mu_k$.)
Similarly, a positive $\lambda_{\chi}$ term tends to shut off the $\chi$
resonance because once the variance of the $\chi$ field becomes large, the
$3\lambda_{\chi} \chi^2$ term acts like an effective mass term,
hence damping the resonance \cite{firstref}.
However in our case we have the extra driving term proportional to
$\Phi_k$, which also depends on $m_{\chi}^2$ and $\lambda_{\chi}$.  Hence
it is not obvious that adding these two parameters will definitely damp
the resonance.  This remains an interesting topic for further research.

\ssec{Escape route 4:  Suppression of Super-Hubble Initial
Conditions}\label{escape4}

The largest resonance effects occur when $q \gg 1$, that is, when the
coupling $g$ is much larger than $(m/M_{\rm pl})^2$.  During inflation,
when $\varphi$ is slowly rolling, such a large coupling $g$ would lead
to a large effective mass for the $\chi$ field, $m_\chi^2 \sim g
\varphi^2 \gg m^2$.  It has been drawn to our attention that this large
effective $m_\chi^2$ during inflation might suppress the value
of the $\chi$ field at the beginning of preheating on $k/a \ll H$
scales, relative to modes in the $k \rightarrow \infty$
limit.\cite{DWpers}
Such a suppression of the amplitude of the $\chi_k$ modes in the $k
\rightarrow 0$ limit might suggest that the large-$k$ modes would have
a \lq\lq head start" at preheating, and would therefore go nonlinear
before the super-Hubble modes would, effectively damping the resonance
via backreaction {\em before} the super-Hubble modes experienced
significant resonant growth.

This situation, however, is complicated by several factors, each of
which deserves further attention:

\begin{itemize}

\item The Floquet indices $\mu_k \rightarrow 0$ rapidly with growing 
$k$.  During the first few inflaton oscillations, for multi-field
models with large $q$, modes grow due to stochastic resonance.  In this
case, the amplification factor goes as \cite{KLS2} $e^{\mu_k} - 1
\simeq 2 \exp [ - \pi k^2 / (2 a^2 m^2 \sqrt{q_{\rm eff}})]$, which
obviously decays rapidly with $k$.  This shows that large-$k$ modes
will not grow at a higher exponential rate during preheating, even 
if they had a large \lq\lq head start."

At later times, when the evolution is closer to ordinary Floquet
theory, the above conclusion remains:  $\mu_k$ will be larger
for the lowest resonance band, at small-$k$, and drops off
rapidly with increasing $k$.

\item Assuming that long-wavelength modes of interest first cross the
Hubble scale 50 $e$-folds before the end of inflation, these modes
would naively be suppressed by a factor of $a^{-3/2} \sim e^{-75} \sim
10^{-30}$.  Even a suppression of $\delta \chi_k (t_0)$ initial
conditions for the small-$k$ modes of order $10^{-30}$ compared with
large-$k$ modes is not necessarily significant when one considers
exponential instabilities at preheating:  $\mu_k/m \sim O(1)$ for $k
\rightarrow 0$ modes when $q \gg 1$, so that small-$k$ modes could {\em
still} go nonlinear first if $\mu_{k \sim 0} \geq 75 \mu_{k \rightarrow
\infty}$.  Considering that $\mu_k$ falls off exponentially with $k$,
this constraint can easily be satisfied over large regions of parameter
space.

\item Not only are the $\mu_k$ {\em much smaller} for large $k$ than
for small $k$, but the {\em widths} of the resonance bands, $\Delta k$, 
are likewise
much narrower for large $k$.  This is crucial since backreaction is
controlled by $k$-space integrals, such as $\Sigma_I \sim \int dk k^2
\vert \delta \varphi_{Ik} \vert^2$, for the field fluctuations $\delta
\varphi_I$.
The decrease in the resonance-band window at large $k$ is partially
compensated for by the larger phase-space volume, $k^2 dk$ factor, and
therefore requires careful quantitative study.

\item This scenario for suppressing the amplitude of super-Hubble modes
neglects the quantum-to-classical transition, which remains very
subtle.  The standard procedure in nonequilibrium quantum field theory
is to {\em define} the initial conditions as $\delta \varphi_{Ik}
(t_0) \equiv  1/\sqrt{2 \Omega_k (t_0)}$ and $\left[d \delta
\varphi_{Ik}/dt \right]_{t_0} \equiv -i \sqrt{\Omega_k
(t_0)/2}$ for {\em all} $k$.\cite{DBDK,son,bhpinit}.
This prescription
has been used in every numerical and nearly every analytical study of
preheating.\cite{num}  Note that these initial conditions
for the start of
preheating suppress large-$k$ modes and favor small-$k$ modes.  By
adopting our scale-invariant initial conditions for $\delta
\varphi_{Ik}$ in our simulations in Sec. \ref{early}, we have thereby
neglected this additional {\em enhancement} of super-Hubble modes.

\end{itemize}

Because of the many subtleties involved in the question of initial
conditions, this obviously warrants further study.  
It is crucial to note, however, that even if super-Hubble modes were
suppressed by a large factor during inflation, some modes $\Phi_k$ are
still 
very likely to go nonlinear during preheating, at some scale $k$:  if
preheating happens to the matter-field modes at any particular scale $k$,
then the metric perturbations $\Phi_k$ {\em must} necessarily experience
similar exponential growth; this is fixed by the constraint equation, Eq. 
(\ref{multi1}). And, as emphasized above in Sec. \ref{sec:Back}, since
preheating only ends via nonlinear mechanisms, the exponential
amplification of these
$\Phi_k$ modes will continue into the nonlinear regime.  Whether or not
the strongest
amplification and earliest transition to nonlinearity occurs on
super-Hubble or sub-Hubble scales, it is bound to happen in {\em some}
region of $k$-space during preheating.

\section{Hierarchical classification of inflationary models}
\l{sec:Hierarchy}

Here we examine a variety of inflationary scenarios for
possible sensitivity to strong preheating. 


${}$

\underline{\em Simple chaotic inflation}

${}$

The start of reheating in chaotic inflation is characterized by very
large values of $\varphi_0
\approx M_{\rm pl}$.\footnote{The exact moment when oscillations
are said to start, and hence the value of the initial amplitude of
oscillations, $\varphi_0$, is
rather subtle. Using the criterion that
it is defined by the time when the first of the slow-roll parameters
$\epsilon$ or $\eta$ become unity, gives $\varphi_0 = \sqrt{2}M_{\rm pl}$ 
for the quadratic potential and $\varphi_0 =
2\sqrt{3} M_{\rm pl}$ for the quartic potential. Using such large
values makes the resonance much stronger. To obtain a lower limit we
use $\varphi_0 \approx 0.3 M_{\rm pl}$, as typical  in the
literature, see e.g. \cite{KLS2,num}.}
In simple polynomial models $V = \lambda \phi^{2n}$, $\lambda$ is
forced to be very small due
to the COBE CMB results, with $\lambda \approx 10^{-12}$ for $n = 2$, or
$m \approx 10^{-6} \> M_{\rm pl}$ for $n = 1$.  Hence we find the result
for scalar fields coupled to the inflaton, as in
Eq. (\ref{qchaotic}),
\beq
q \sim g \times 10^{10} ,
\l{eq:qchaotic}
\eeq
so that for natural couplings $g \in [10^{-6},10^{-2}]$ we have $q \in
[10^{4},10^{8}]$. In other words, large resonance factors are generic in
the simplest models of chaotic inflation, when scalar fields in addition
to the inflaton are included.

A very important point to notice is that the large $q$ arise due to the
confluence of the largeness of $\varphi_0$ and the smallness of the
inflaton mass {\em during reheating}. If either of these is given up, the
resonances can be expected to be naturally much weaker.

Hence, in a chaotic inflation model based on a potential  which exhibits a
large change in the inflaton effective mass between the inflationary phase
and the reheating phase, the $q$ parameter might easily be several orders
of magnitude smaller due to the strong curvature of the potential at
reheating. 

Finally, in the case where the potential exhibits self-interaction
for the inflaton, e.g. $V = \lambda \phi^4/4$, it has been strongly
argued that the value $\lambda \sim 10^{-12}$ is unnecessarily small,
arising from a naive identification of classical perturbations
(responsible for the CMB anisotropies) with quantum fluctuations, and
that a value nearer $\lambda \sim 10^{-6}$ is more accurate
\cite{CH95}.  If this is true, then for models with a {\em massless}
quartically self-coupled inflaton coupled to {\em distinct} bosons,
the resonance parameter $q$ would be reduced by around $10^6$, since
the square of effective frequency of the inflaton's oscillations,
$\omega^2 \propto \lambda \varphi^2_0$, appears in the denominator
for $q$.  Naturally, this would be a very large change, which, for
moderate couplings $g < 10^{-3}$, would move the system into the mild
or weak resonance regimes.  This change in $\lambda$, however, would
not necessarily weaken the resonances for chaotic inflation models
with a massive inflaton coupled to distinct fields.

${}$

\underline{\em Hybrid and  multiple inflation}

${}$

General hybrid models can exhibit strong preheating with large $q$
parameters \cite{GBL98}, and to this extent they will suffer strong
distortion of the
metric perturbation spectrum during any oscillatory phase. What would be
interesting would be to study models in which an oscillatory phase with
large resonance parameters occurs between successive phases of
inflation. Once the oscillations have finished, the amplified
spectrum would be smoothed and stretched during the second
inflationary phase leaving a
highly non-trivial imprint on the CMB, assuming the second phase did not
last too long.

The extra freedom in these models implies that it should be
possible to construct models with very similar features to those of the
broken spectral index genre, which are able to fit current CMB and large
scale structure data better than standard inflationary models
\cite{bsi,subir}.

\begin{figure}
\epsfxsize=3.4in
\epsffile{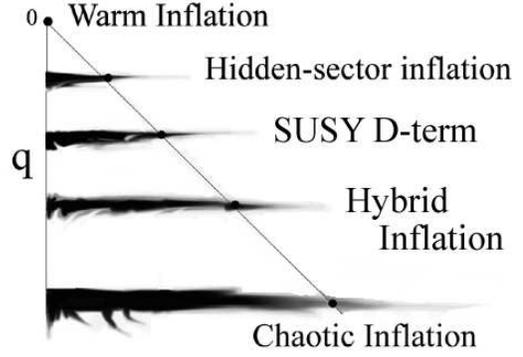}
\caption{A schematic representation of the hierarchy of inflationary
models with regard to their strength of preheating. At the top,
where $q = 0$, sits warm inflation. At the bottom, where
typically $q \gg 1$, sit the simplest chaotic inflation models. Most
other inflationary models lie in between and exhibit the
effects discussed in this paper to varying degrees depending on
precise details of the couplings between the inflaton and other fields. }
\l{fig:hiermodel} \end{figure}

${}$

\underline{\em Supersymmetric (SUSY) inflation}

${}$

Consider perhaps the simplest SUSY theory based on the superpotential
\cite{gia}:
\beq
W = \alpha S \overline{\varphi} \varphi - \mu^2 S .
\eeq
This is the most general form consistent with R-parity under which the
inflaton singlet, $S$, transforms as $S \rightarrow e^{i\theta}S$, the
superpotential transforms as $W
\rightarrow e^{i\theta}
W$, and the product $\overline{\varphi} \varphi$ is invariant. The
unbroken
supersymmetric potential corresponding to this superpotential is:
\beqn
V &=& \alpha^2 |S^2|(|\varphi|^2 + |\overline{\varphi}|^2)
\nonumber\\
&&~~~ + |\alpha \varphi
\overline{\varphi} - \mu^2|^2 + \mbox{$D$-terms}\,.
\eeqn
The $D$-terms vanish along the $D$-flat direction $|\varphi| =
|\overline{\varphi}^*|$. Inflation occurs for $S \gg \mu/\sqrt{\alpha}$
at the minimum of the potential ($\langle \varphi\rangle =
\langle \overline{\varphi} \rangle = 0$), which is not supersymmetric,
with $V =
\mu^4$. Quantum corrections break the flatness of the potential and cause
slow-roll but as long as supersymmetry is broken softly the induced
curvature will be small and the inflaton  very light, as in chaotic
inflation above. If this were to remain the situation, these theories
would have the same problems as the simple chaotic inflation models.
However, for $S \leq \mu/\sqrt{\alpha}$, $V$ has a new minimum which is
supersymmetric ($V = 0$) corresponding to $\langle S \rangle = 0,
\langle \varphi \rangle = \mu/\sqrt{\alpha}$ \cite{gia}. Reheating
is expected to
complete during oscillations about this new minimum. The amplitude of both
$S$ and $\varphi$ oscillations is of order $\mu/\sqrt{\alpha}$, while
their
masses are of order $m_{S}^2 \approx 4 \mu^2$ and $m_{\varphi}^2 \approx
8\alpha \mu^2$. Hence, assuming no
other fields, the resonance parameters at the final oscillation stage are:
\beq
q_{S} \approx \frac{\alpha^2  \mu^2}{4 \mu^2 \alpha}  \sim \alpha \sim
10^{-2}
\eeq
\beq
q_{\varphi} \approx \frac{\alpha^2  \mu^2}{8\mu^2 \alpha^2} \sim 0.1 ,
\eeq
since $\alpha \sim 10^{-2}$ to fit observations \cite{gia}. Hence,
in this most minimal of supersymmetric models, there is little resonance
in the primary oscillation phase and in the inflaton and Higgs superfields
$\varphi$. In this sense it is closer to the single-field models since it
contains only one energy scale ($\mu^2$) and a single dimensionless
coupling, $\alpha$, which is forced to be small.

However, if we couple for example, another scalar field $\chi$ to
$S$ and $\varphi$, with coupling $g^2 \gg \alpha^2$,
then  the  resonance parameters would of the order
$g^2/\alpha^2 \sim g^2 \times 10^{4}$, so that if $g$ were near
strong coupling, the strong resonances we have discussed here would
occur. However, since these extra couplings would typically give
mass to other fields, (e.g.  the heavy right-handed neutrinos of the
theory \cite{LSS96}) and would play a role in determing the
baryon asymmetry, building a fully consistent model with $g^2 \sim 1$ 
appears technically very difficult.

${}$

\underline{\em Hidden sector inflation}

${}$

In these string-inspired  models \cite{subir}, the inflaton in the hidden
sector only has gravitational couplings to the fields of the visible
sector and hence the $q$ parameter is very small, being of order
unity, as occurs for gravitational waves \cite{gw}. The amplification
of metric perturbations is essentially negligible in these models, as
it is in the quintessential inflation model of Vilenkin and Peebles
\cite{VP98}.

${}$

\underline{\em Warm inflation}

${}$

At the opposite end of the scale from simple chaotic inflation models is
warm inflation and its variants \cite{arjun}, in which inflation
occurs due to strong overdamping of the inflaton due to couplings to
other fields. Reheating does not occur due to coherent oscillations
but rather occurs continuously during inflation with
natural exit to a radiation-dominated FRW universe. Since there are no
coherent oscillations to speak of, $q = 0$, and there is no resonant
amplification of metric perturbations at all.

\section{Conclusions and future issues} \l{sec:Conc}

This paper continues the exploration of metric perturbations'
evolution during the post-inflationary epoch known as preheating. In
spirit it is a continuation and development of the study in Paper I
\cite{BKM1}, but with significant advances in conceptual and
numerical analysis of the problem.  We have identified a number of
important issues that distinguish the evolution of metric
perturbations during the violent preheating epoch ($q \gg 1$) from
metric evolution in old theories of reheating.  Primary among these
is the resonant amplification of {\em both} super- and sub-horizon
modes of $\Phi$, the gauge-invariant scalar metric perturbation. 

We demonstrate explicitly that such amplification {\em does
not} violate causality -- it is merely a reflection of the perfection
of the coherence of the inflaton condensate at the very end of
inflation, which is correlated on vastly super-Hubble scales in
standard models of inflation.

Such resonances depend heavily upon the presence of
multiple matter-fields coupled via non-gravitational interactions.
In the multi-field case,
entropy perturbations are generic, and although they may be small during
inflation, they are
resonantly amplified during strong preheating.  This leads to the
destruction of the conservation
of traditional quantities such as the Bardeen parameter $\zeta$,
traditionally used to transfer the metric perturbation $\Phi$ from
inflation to photon decoupling. This implies that direct numerical
integration of the governing equations must be used.

The wide resonances at $q \gg 1$ destroy the scale-invariance of the
inflationary power spectrum and could hugely over-produce large-angle
anisotropies in the Cosmic Microwave Background if left undamped after
preheating. The resonances can be so strong as to force the
perturbations to go nonlinear, requiring study of the full,
unlinearized Einstein
field equations (see \cite{PE} for an exciting  start in this direction in
the single-field case). One issue that is likely to be very
important in understanding the full extent of amplification of the
super-Hubble modes is the {\em break-up} of the inflaton
condensate. As second-order effects become important, we expect on
general grounds that the perfect coherence of the oscillations of the
inflaton condensate will be lost, and the coherence length of the
homogeneous mode will be reduced. This will tend to shut off the
super-Hubble resonances, following essentially from
the induced super-Hubble inhomogeneity.

We wish to emphasize that this powerful
growth in the metric perturbations is robust and cannot be swept
away or removed by fine-tuning the initial value of $\Phi$ during
inflation. The primary reason why fine-tuning fails is that
the resonances of preheating end typically only through
backreaction effects, which require at least study of second-order
metric perturbation effects and a thorough understanding of
mode-mode coupling. This is the first time inflationary models
have been faced with this complication, bringing them nearer in
spirit to the nonlinear field theories used in topological defect
studies. This mode-mode coupling tends to re-establish the
scale-invariance of the power spectrum, but this time with a
spectral index which is determined by the nonlinear structure of
the Einstein field equations, in analogy with the situation with
turbulence and the Navier-Stokes equations. Memory of the initial
spectral index is lost and the perturbations tend to an active,
incoherent state, again  mimicking topological defects. This may have
profound implications for the small-scale CMB and in particular
for the secondary Doppler peaks, as a means for distinguishing between
inflation
and defect theories, assuming the mode-mode coupling is preserved
on relevant scales until decoupling.

As a means for quantifying the amplification of perturbations, we
have introduced here the concept of the {\em time to nonlinearity},
$t_{\rm nl}$, which provides  a robust (though non-unique) and useful
generalization of the Floquet index $\mu_k$ for expanding universes.
It quantifies the time at which the linearized equations break
down.

The metric preheating we have developed here, despite being based on a
simple model, yields very strong deviations from the old theory of
perturbative evolution. These deviations depend crucially on a number of
conceptual points, highlighted in Sec. \ref{sec:Concept}, related to
the existence of multiple, non-gravitationally interacting
fields at preheating. While certain regions of preheating
$q$-parameter space appear to be observationally ruled out by the
CMB, there appear a number of interesting and even exciting
mechanisms for making
the strong resonances acceptable. Secondary phases of inflation
are expected to damp the amplified modes and stretch them. This
allows one to put more freedom into the original spectrum while
bringing the overall amplitude down to an  observationally
acceptable level.

Alternatively, the resonance may be significantly weaker if the
fields into which the inflaton is decaying have strong
self-interaction, or are fermionic. The decay into fermions is
particularly interesting since it is constrained by the Pauli
exclusion principle, which limits the amplification of $\Phi$.
While we have discussed the basic mechanisms at work, both of these
aspects deserve closer attention.

For the theory of preheating itself, one of the implications of
studying the complete system of equations including metric
perturbations is the appearance of rescattering -- the production
of quanta due to couplings between the fluctuations of different
fields -- even {\em at linear order}. Previously this had been
thought \cite{KLS2} to be a purely nonlinear process.
On smaller scales, the amplification of scalar metric
perturbations strongly enhances primordial black hole formation
and gravitational wave emission via rescattering.

We note that all inflationary models are
definitely not born equal with regard to these resonances. Simple
chaotic inflation models, with their large amplitude oscillations
and small effective mass, yield extremely strong resonances.
Alternatively, models with big changes to the effective mass
and/or small amplitude oscillations or weak couplings between the
inflaton and other fields, exhibit much weaker resonances.

To conclude, we believe it is important to analyze the limitations of
the work presented here. Since preheating is a problem involving
multi-field, non-equilibrium, semi-classical quantum gravity on a
dynamic background, it is clear that the essentially classical model
we have used is not ideal. In addition, we have uncovered a serious
ambiguity related to the definition of the homogeneous components of
the non-inflaton fields. The standard definition in the metric 
perturbation
literature ignores the contribution of $\Phi(k = 0)$ to the evolution
of homogenous background quantities. Since the energy in this
mode grows very rapidly, we have argued that its
influence on the background should be included as a first step to
studying the full backreaction problem. 

Other aspects of preheating that should be addressed in
future work include:

(i) Can one formulate a consistent  quantum and
non-equilibrium framework which includes the metric perturbations? 
In particular, does the Schwinger-Keldysh closed-time-path (CTP)
formalism \cite{DBDK,f1,CH94} carry through to this more complex
setting?  Further, can one reconcile the standard relativistic
definitions of background quantities with the order-parameter
\cite{CH95} and other field-theoretic definitions? 

(ii) Backreaction issues: can one go beyond the simple linearized
equations in a realistic, analytical manner? Is there a
relatively simple and consistent Hartree or large-$N$ approximation
for Eqs. (\ref{multi1a}-\ref{multi1})?

(iii) The transfer function to decoupling and thermalization: how
does thermalization proceed and how do metric perturbations evolve
through this phase? How does the viscosity arising from
finite temperature effects alter perturbation evolution? 
\cite{BM,CG97,NRS,ZPM}

(iv) Quantum gravity corrections: stochastic backreaction effects are
expected to be strong during preheating. This implies a
stochastic evolution of background quantities such as the scale
factor and zero modes \cite{CH94}. The evolution of the metric is
determined by a Langevin-type equation which describes the
non-equilibrium dynamics of the gravitational field.  The effects of
these stochastic fluctuations have already been analyzed
phenomenologically in the
context of preheating without metric perturbations
\cite{noise,bass98,BT98}, and found generically to enhance particle
production. We have similar expectations on very general grounds for
the case in which metric fluctuations are included.  (See also
\cite{hushio}.)  Almost surely
going to the (white-noise) stochastic limit will not help reduce the
amplification of metric perturbations, though it will change the 
wavelength-dependence of the amplification.


\[ \]
{\bf Acknowledgments:}

The authors have benefited greatly from discussion with a number of
people. We thank John Barrow, Arjun Berera, Robert Brandenberger,
Marco Bruni, Richard Easther, Fabio Finelli, Larry Ford, Alan Guth, Rachel
Jeannerot,
Janna Levin, Andrew Liddle, Matthew Parry, Subir Sarkar, Dennis Sciama,
Dam Son, David Wands and members of the Stimulated Emission group for
these discussions.  DK acknowledges partial support from NSF grant
PHY-98-02709. 


\end{document}